\documentclass[prd,showpacs,superscriptaddress,preprintnumbers,amsmath,amssymb,nofootinbib]{revtex4}

\usepackage{graphicx}
\usepackage{dcolumn}
\usepackage{bm}
\usepackage{color}
\usepackage{epsfig}
\usepackage{amssymb}
\usepackage{amsmath}

\usepackage[usenames, dvipsnames]{xcolor}
\usepackage{url}
\usepackage[linktocpage=true]{hyperref}
\hypersetup{
   colorlinks   = true,
    linkcolor= Blue,
    citecolor= Blue,
    urlcolor= Blue}

\newcommand{\MS}{\ensuremath{\overline{\text{MS}}}}
\newcommand{\bea}{\begin{eqnarray}}
\newcommand{\eea}{\end{eqnarray}}
\newcommand{\nn}{\nonumber}

\newcommand{\eq}[1]{Eq.~(\ref{#1})}

\newcommand{\be}{\begin{equation}}
\newcommand{\ee}{\end{equation}}

\begin{document}

\title{\boldmath \Large \bf Improved analysis of the (${\mathcal O}_7$,\,${\mathcal O}_7$) contribution to $\bar{B} \rightarrow X_s\gamma \gamma $ at $O(\alpha_s)$}

\author{H. M. Asatrian}
\email[Electronic address:~]{hrachia@itp.unibe.ch}
\affiliation{Yerevan Physics Institute, 0036 Yerevan, Armenia}
\author{C. Greub}
\email[Electronic address:~]{greub@itp.unibe.ch}
\affiliation{Albert Einstein Center for Fundamental Physics, Institute
  for Theoretical Physics,\\ University of Bern, CH-3012 Bern,
  Switzerland}
\author{A. Kokulu}
\email[Electronic address:~]{akokulu@liverpool.ac.uk}
 \affiliation{Department of Mathematical Sciences, University of Liverpool,\\ L69 7ZL Liverpool, United Kingdom} 
 %


\begin{abstract}
\vspace{1cm}
The present study is devoted for an improved analysis of the
self-interference contribution of the electromagnetic dipole operator ${\cal O}_7$
to the double differential decay width $d\Gamma/(ds_1 ds_2)$ for the inclusive
$\bar{B} \to X_s \gamma \gamma$ process, where the kinematical variables $s_1$
and $s_2$ are defined as $s_i=(p_b - q_i)^2/m_b^2$ with $p_b$, $q_1$, $q_2$ 
being the momenta of the $b$-quark and two photons. This calculation completes
the NLL QCD prediction of the numerically important self-interference
contribution of ${\mathcal O}_7$ by keeping the full dependence on the strange-quark
mass $m_s$, which is  introduced to control possible collinear
configurations of one of the photons with the strange
quark. Our results are given for exact $m_s$, in contrast to an earlier work
where only logarithmic and constant terms in $m_s$ were retained. This improved NLL result for
the (${\mathcal O}_7$,\,${\mathcal O}_7$)-interference contribution shows that finite $m_s$ effects
are only sizable near the kinematical endpoints of the spectrum
$d\Gamma/(ds_1ds_2)$. At the level of the branching ratio, in the
phase-space region considered in this paper, it is observed that
$\rm Br[\bar{B}\to X_s \gamma \gamma]$ does not develop a sizable $m_s$ dependence:
the impact on this branching ratio is less than $5\%$ when $m_s$ is varied between $400$--$600$
MeV. For the same phase-space region finite strange quark mass effects for
the branching ratio  are less than $7\%$.

\end{abstract}


\preprint{LTH 1103}

\maketitle

\section{Introduction}
\label{sec:Introduction}

In the Standard Model (SM), flavor changing neutral current transitions (such
as $b\to s\gamma(\gamma)$) are suppressed since they are loop-induced. 
When going beyond the SM, such processes could provide a unique
source for probing physics indirectly at the TeV scale. For
instance, in the two-Higgs-doublet-model (2HDM) of type II 
the inclusive singly radiative decay $\bar{B} \rightarrow X_s\gamma$
is known to have provided a very stringent (almost $\tan\beta$-independent) lower
bound on the charged Higgs boson mass to be $m_{H^{\pm}}>480$
GeV at $95\%$ CL. This limit has been 
obtained by comparing the recent
experimental data for $\rm Br[\bar{B} \rightarrow X_s\gamma]$ with the corresponding theoretical 2HDM results, which are based
on the next-to-next-to-leading-logarithmic (NNLL)
SM results \cite{Misiak:2015xwa}, as well as on the NLL \cite{Borzumati:1998tg,Borzumati:1998nx,Ciuchini:1997xe}
and NNLL \cite{Hermann:2012fc} charged Higgs contributions to the various Wilson coefficients.

Although the branching ratio for the singly radiative decay $\bar{B} \to X_s
\gamma$ is much larger, the double radiative decay
$\bar{B} \to X_s \gamma \gamma$ possesses certain advantages.
In contrast to the singly radiative decay, 
the current-current operators ${\mathcal O}_{1,2}$ contribute to the double
radiative decay at order $\alpha_s^0$ precision (through one-particle irreducible
one-loop diagrams), leading to an interesting interference pattern with the
contributions associated with the electromagnetic dipole operator  ${\mathcal
  O}_{7}$ already at LL precision. As a result, potential new physics should be
clearly visible not only in the total branching ratio, but also in the
differential distributions.

The process $\bar{B} \to X_s \gamma \gamma$ is of direct interest to the new Belle II experiment (SuperKEKB)
in Japan \cite{superKEKB:online}, which aims to detect branching ratios as
small as $10^{-8}$ or smaller and will start taking $B$ data in 2018
\cite{HerediadelaCruz:2016yqi,Aushev:2010bq}. This calls for more precise SM
calculations of this observable. The status of the related works can be
summarized as follows: The SM estimates of the branching ratios for $\bar{B}
\to X_s \gamma$ \cite{Misiak:2015xwa,Misiak:2006zs}
and $\bar{B} \to X_s \ell^+
\ell^-$ are now available even to NNLL precision (see e.g.
\cite{Hurth:2010tk,Buras:2011we} for reviews). Regarding the $\bar{B} \to X_s \gamma
\gamma$ decay, the leading logarithmic
(LL) prediction for the branching ratio was known since a long
time \cite{Simma:1990nr,Reina:1996up,Reina:1997my,Cao:2001uj}, while the
first attempts towards a NLL calculation were only made years later by us
\cite{Asatrian:2011ta,Asatrian:2014mwa}, in which the QCD corrections to
the numerically dominant
(${\cal O}_7$,~${\cal O}_7$) contribution were worked out in certain
approximations which will be detailed in the next paragraph. In 2015, we also provided the contributions
stemming from the self-interference of chromomagnetic dipole
operator ${\cal O}_8$ \cite{Asatrian:2015wxx}. 

In Ref. \cite{Asatrian:2011ta} we calculated order $\alpha_s$ corrections
based on the operator ${\cal O}_7$ to the double differential decay
width $d\Gamma/(ds_1 \, ds_2)$, by taking into account only the leading power
terms in $s_3$ in the underlying triple differential decay width 
$d\Gamma/(ds_1 \, ds_2 \, ds_3)$, where $s_3$ is the normalized invariant
mass squared of the hadronic particles in the final state.  In
Ref. \cite{Asatrian:2014mwa}, we worked out the double differential 
decay width based on the triple differential width, retaining all powers
w.r.t. $s_3$. In that work we approximated, however, the dependence on the
strange quark mass $m_s$, by only keeping logarithmic terms in $m_s$ and
the terms which are independent of $m_s$. As $m_s$ is interpreted as a
constituent mass in our set-up, varying in a range between 400 and 600
MeV, this approximation is somewhat questionable. In the present work,
we therefore give NLL results which take into account the full $m_s$
dependence. 
We denote these results as {\it exact results} and the
results of  Ref. \cite{Asatrian:2014mwa} as {\it results in the limit $m_s \to 0$}.

We should mention that this inclusive process has
also been analyzed in some new physics scenarios
\cite{Reina:1996up,Cao:2001uj,Gemintern:2004bw}. Also, spectator quark and
long distance (resonant) effects were studied in the literature (see
e.g. \cite{Chang:1997fs,Hiller:2004wc,Ignatiev:2003qm}, \cite{Reina:1997my}
and references therein). Further, there have also been several studies on the
corresponding exclusive channels $B_s \to \gamma \gamma$ and $B\to K \gamma
\gamma$, both within
\cite{Reina:1997my,Chang:1997fs,Hiller:1997ie,Bosch:2002bv,Bosch:2002bw,Hiller:2004wc,Hiller:2005ga,Lin:1989vj,Herrlich:1991bq,Choudhury:2002yu}
and beyond the SM
\cite{Gemintern:2004bw,Hiller:2004wc,Hiller:2005ga,Aliev:1997uz,Bertolini:1998hp,Bigi:2006vc,Devidze:1998hy,Aliev:1993ea,Xiao:2003jn,Huo:2003cj,Chen:2011te,Qin:2009zzb}.

Our paper is structured as follows.
In section \ref{sec:framework} we discuss the theoretical framework and some
preliminaries for the calculations.
In section \ref{sec:leadingorder} we work out the double differential distribution 
$d\Gamma_{77}/(ds_1 ds_2)$ in leading order, i.e. without taking into
account QCD corrections to the matrix element 
$\langle s \gamma \gamma |{\cal  O}_7|b\rangle$. 
In this section we also give the order $\alpha_s^0$ results when
including the effects of the operators ${\cal O}_1$ and ${\cal O}_2$, keeping
the full dependence on $m_s$.

In section \ref{sec:NLL} we calculate virtual and bremsstrahlung QCD
corrections to the double differential decay width $d\Gamma_{77}[\bar{B} \rightarrow X_s\gamma \gamma]/(ds_1 \, ds_2)$ associated with the operator ${\mathcal O}_7$, 
keeping the full dependence on the strange-quark mass $m_s$. 
In section \ref{sec:numerics} we give numerical illustrations of our
results. In section \ref{sec:summary} we give a brief summary of our findings.

\section{Theoretical framework and kinematical cuts}
\label{sec:framework}

We begin our calculation by identifying the effective Hamiltonian governing
$b\to s\gamma(\gamma)$ transition, after integrating out the heavy degrees of
freedom in the SM. This Hamiltonian reads
\be
 {\cal H}_{eff} = - \frac{4 G_F}{\sqrt{2}} \,V_{ts}^\star V_{tb} 
   \sum_{i=1}^8 C_i(\mu) {\cal O}_i(\mu)  \, ,
\label{Heff}
\ee
where the operators are defined according to \cite{Chetyrkin:1996vx} as:
\vspace{0.1cm}
\be
\begin{array}{llll}
{\cal O}_1 \, &\!= 
 (\bar{s}_L \gamma_\mu T^a c_L)\, 
 (\bar{c}_L \gamma^\mu T_a b_L)   \,,  &\quad  
{\cal O}_2 \, &= 
 (\bar{s}_L \gamma_\mu c_L)\, 
 (\bar{c}_L \gamma^\mu b_L)\,,   \\[1.002ex]
{\cal O}_3 \, &\! = 
 (\bar{s}_L \gamma_\mu b_L) 
 \sum_q
 (\bar{q} \gamma^\mu q)   \,, &\quad  
 {\cal O}_4 \, &= 
 (\bar{s}_L \gamma_\mu T^a b_L) 
 \sum_q
 (\bar{q} \gamma^\mu T_a q)\,,  \\[1.002ex]
{\cal O}_5 \, &\! = 
 (\bar{s}_L \gamma_\mu \gamma_\nu \gamma_\rho b_L) 
 \sum_q
 (\bar{q} \gamma^\mu \gamma^\nu \gamma^\rho q)   \,,  &\quad
{\cal O}_6 \, &=
 (\bar{s}_L \gamma_\mu \gamma_\nu \gamma_\rho T^a b_L) 
 \sum_q
 (\bar{q} \gamma^\mu \gamma^\nu \gamma^\rho T_a q)\,,  \\[1.002ex]
{\cal O}_7 \, &\! = 
  \frac{e}{16\pi^2} \,\left[     
 \bar{s} \sigma^{\mu\nu}   \left( \bar{m}_b(\mu) \,R\,   +  \, \bar{m}_s(\mu) \,L  \right)  F_{\mu\nu} b  \right]  \,, &\quad 
{\cal O}_8 \, & = 
  \frac{g_{s}}{16\pi^2} \,\left[     
 \bar{s} \sigma^{\mu\nu}   \left( \bar{m}_b(\mu) \,R\,   +  \, \bar{m}_s(\mu) \,L  \right)  T^a G^a_{\mu\nu} b  \right]  \, .
\end{array} 
\label{opbasis}
\vspace{0.2cm}
\ee
In \eq{opbasis}, $T^a$ ($a=1,8$) are the $SU(3)$ color generators, $e$ and
$g_s$ are the electromagnetic and the strong couplings and $\bar{m}_s(\mu)$
and $\bar{m}_b(\mu)$ are the running $s$ and $b$-quark masses defined in the
$\MS$-scheme. Note that we keep the term involving $m_s$ in the operator
${\cal O}_7$, as we keep the full dependence on $m_s$ in our work. Further, 
\eq{Heff} takes this compact form only after neglecting
the small $V_{ub} V_{us}^*$ element (as $V_{ub} V_{us}^* \ll V_{tb} V_{ts}^*
$) and exploiting the unitarity of the unitarity of
Cabibbo--Kobayashi--Maskawa (CKM) matrix.

In the effective theory framework, the calculation for the branching
ratio (or for a specific differential distribution) can be divided into two
parts. The first one deals with perturbative matching calculations of the Wilson
coefficients $C_i(\mu)$ appearing in \eq{Heff} at the large scale ($\mu \sim
m_W$), followed by solving the renormalization-group-equations (RGE) for
these coefficients\footnote{Solving RGEs requires computing
  anomalous-dimension-matrices (ADM) of the effective operators to the desired
  order (see e.g. Refs.~\cite{Gorbahn:2004my,Gorbahn:2005sa,Czakon:2006ss} for
  the impressive three and four-loop ADM contributions).} to obtain their
values at the decay scale ($\mu \sim m_b$). The second part consists of calculating
the matrix elements of the operators in \eq{opbasis}. At the bottom scale,
the strong coupling is still small enough ($\alpha_s(m_b)\sim 0.22$) such that
perturbative calculations of the matrix elements are possible. 

For the process of
interest, the Wilson coefficients at the low scale $C_i(\mu \sim m_b)$ are
available today even to NNLO precision (see e.g. the reviews
\cite{Hurth:2010tk,Buras:2011we} and references therein). On the other hand,
the matrix elements $\langle s \gamma \gamma|{\cal  O}_i|b\rangle$ and 
$\langle s g \gamma \gamma |{\cal  O}_i|b\rangle$, which in a NLL calculation are needed to order
$g_s^2$ and $g_s$, respectively, are only partially known by now (see
\cite{Asatrian:2011ta,Asatrian:2014mwa,Asatrian:2015wxx} for the details of the provided
NLL contributions and \cite{kokulu:2016forbelle2writeup} for a recent summary).

In the present paper, we calculate $O(\alpha_s)$ corrections arising from the
self-interference contribution of the electromagnetic dipole operator ${\cal
  O}_7$ to the double differential decay width $d\Gamma/(ds_1 ds_2)$ for
$\bar{B} \to X_s \gamma \gamma$, where the kinematical variables $s_1$ and
$s_2$ are defined as $s_i=(p_b - q_i)^2/m_b^2$ with $p_b$, $q_1$, $q_2$ being
the four-momenta of the $b$-quark and two photons. At this order in
$\alpha_s$, this involves contributions with three (virtual) and four
particles (bremsstrahlung) in the final state. The key difference to our study
in Ref.~\cite{Asatrian:2014mwa} is that we keep the
{\it full} dependence on strange-quark mass in our results.

Kinematically, the $(s_1,s_2)$-region accessible to the three body decay $b\to
s\gamma\gamma$
is given by (see \cite{Asatrian:2012tp}) 
by\footnote{The phase-space region corresponding to real gluon radiation $b\to s g \gamma\gamma$ is wider than
this. Nevertheless we consider the bremsstrahlung process only in the restricted
region, which is also accessible to the three body decay $b\to s \gamma\gamma$.}
\be
s_1>x_4,\; s_2>x_4,\; 1-s_1-s_2+x_4 > 0 ~;{}~ s_1 s_2 > x_4    \, ,
\label{eq:ps1a}
\ee
where $x_4=m_s^2/m_b^2$. 
In the rest frame of the decaying $b$-quark, one has
a simple relation between $s_i$ variables and  final state photon energies
$E_i$: $s_i=1-2E_i/m_b$. At this stage, we need to impose some kinematical
cuts. First, in order for observing two hard photons, the $s_i$ variables should
be smaller than one. Also, detection of two distinct photons requires
kinematically that their invariant mass is different from zero. It is possible
to satisfy all these requirements using a single physical cut-off parameter $c$
($c > x_4$), by demanding\footnote{In terms of the four particle final state,
  the invariant mass squared $s = (q_1 + q_2)^2/{m_b}^2$ of the two photons can be
  written as $s =1-s_1 - s_2 + s_3$, where $s_3$ is the normalized hadronic
  mass squared. Then, choosing $1-s_1-s_2 > c$, still prevents the photons
  from flying parallel to each other.}
\be
 1-s_1-s_2 > c ~;{}~ ( s_1-c)\, (s_2-c) > c    \, .
\label{eq:ps2cuts}
\ee
Note that the region defined in Eq. (\ref{eq:ps2cuts}) is a subregion of the one
specified in Eq. (\ref{eq:ps1a}).

With these cuts at hand, soft photon related singularities are absent,
whereas there exist kinematical configurations where one of the photons can become
collinear to the $s$-quark. Working with a finite strange quark mass our final
NLL result involves single logarithms of the form
$\log(m_s/m_b)$, whose origin is entirely
related to collinear photon emissions from $s$-quark and {\it not} to
gluons. The reason for this is that QCD-wise, our observable (the double or
triple differential decay width)
is fully inclusive and therefore nonsingular. As a result, all soft and/or
collinear gluon related singularities cancel out in our final result after
adding the corresponding virtual and bremsstrahlung corrections, as a
consequence of the Kinoshita-Lee-Nauenberg (KLN) theorem. However, QED-wise
our observable is not fully inclusive, because we want to observe exactly two
photons in the final state; therefore $\log(m_s/m_b)$ terms remain.

We note that $m_s$, which is initially introduced as an infrared/collinear regulator,
is eventually interpreted to be a mass of constituent type varying between $400-600$
MeV in the final numerics. We believe that this range covers the
non-perturbative uncertainties due to the hadronic substructure of
photons. This approach has also been adopted previously, e.g. by Kaminski {\it
  et al.} in \cite{Kaminski:2012eb} and Asatrian {\it et al.} in
\cite{Asatrian:2013raa,Asatrian:2014mwa,Asatrian:2015wxx}. The experience
gained in these references shows that the constituent mass approach gives
results which are similar to those when using fragmentation functions
\cite{Asatrian:2013raa}. Therefore, we believe that this method is sufficient
to obtain an estimate of the calculated contribution. While the fragmentation
approach seems better from the theoretical point of view, it is not clear that
it leads to better final results in practice, because  the fragmentation
functions (for $s\to\gamma$ or $g\to\gamma$) suffer from experimental
uncertainties, as pointed out in \cite{Asatrian:2013raa}. An alternative could
be to look at the version with ``isolated photons'' a la Frixione
\cite{Frixione:1998jh} which corresponds, however, to a slightly different
observable. Such an approach is beyond the scope of the present paper and is
left for future studies.

\section{Improved leading order results}\label{sec:leadingorder}
\label{sec:LO}
\begin{center}
\begin{figure}[h]
\centering{
\includegraphics[width=0.55\textwidth]{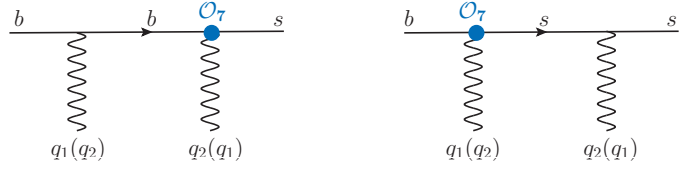}
\caption{Tree-level amplitudes representing the (${\mathcal O}_7$,\,${\mathcal O}_7$) contribution to $b\to s\gamma\gamma$. The symmetric diagrams are understood to be obtained from those shown by interchanging $q_1$ with $q_2$ as indicated in brackets.}      
\label{fig:TreeDiags}
}
\end{figure}
\end{center}

In this section we give the double differential decay width
$d\Gamma_{}/(ds_1 ds_2)$ at lowest order in QCD, keeping the full dependence
on the strange quark mass. 
We define the dimensionless variables $s_1$ and $s_2$  as
\be
s_1=\frac{(p_b-q_1)^2}{m_b^2} \quad ; \quad
s_2=\frac{(p_b-q_2)^2}{m_b^2} \, .
\label{kinematicvariables}
\ee
At lowest order the double differential decay width 
$d\Gamma_{77}^{0}/(ds_1 ds_2)$
is based on the 
diagrams shown in Fig.~\ref{fig:TreeDiags}.
Since the lowest order decay width will also be needed for the
UV-renormalization of the virtual corrections, we present the leading-order
result in $d=4-2\epsilon$ dimensions, keeping terms up to $\epsilon^1$ order in
the expansion. Using $x_4=m_s^2/m_b^2$, we obtain
\begin{eqnarray}
&&\frac{d\Gamma_{77}^{(0,d)}}{ds_1 \, ds_2} = \frac{\alpha^2 \,
     \, m_b^3 
\, |C_{7,eff}(\mu)|^2 \, G_F^2 \,
  |V_{tb} V_{ts}^*|^2 \,  Q_d^2}{1024 \, \pi^5} \, \left(
  \frac{\mu}{m_b}\right)^{4\epsilon}   {\tilde r} \, ,
\label{eq:tree77A}
\end{eqnarray}
with
\begin{eqnarray}
&& {\tilde r}=\frac{\left[ \bar{m}_b^2(\mu) {\tilde r}_a+
 \sqrt{x_4}\bar{m}_s(\mu) \bar{m}_b(\mu) {\tilde r}_b+
 \bar{m}_s^2(\mu) {\tilde r}_c+
\epsilon m_b^2 ( {\tilde r}_1+ {\tilde r}_2+ {\tilde r}_3) \right]
   \left(1-s_1-s_2 + x_4 \right)}{\left(1-s_1\right)^2
   (s_1-x_4)^2 \left(1-s_2\right)^2 (s_2-x_4)^2} \, .
\label{eq:tree77B}  
\end{eqnarray}
Note that the terms proportional to $\epsilon$ in Eq. (\ref{eq:tree77B}) 
will only be involved in the renormalization procedure of the virtual
corrections, leading there to terms of order $\alpha_s$. As in such terms
the running masses $\bar{m}_b(\mu)$ and $\bar{m}_s(\mu)$ can be identified with
the pole masses $m_b$ and $m_s$, we immediately performed this 
identification in the corresponding terms of Eq. (\ref{eq:tree77B}).

\begin{figure}[h]
\begin{center}
\includegraphics[width=0.42\textwidth]{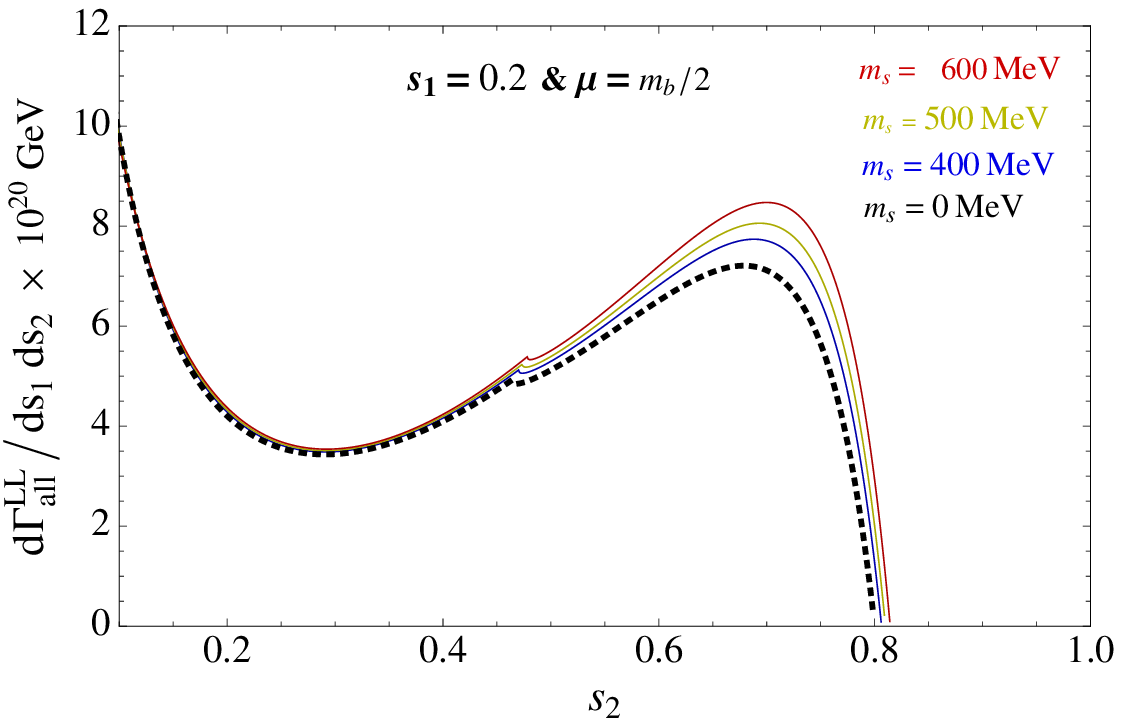}
~~~\includegraphics[width=0.42\textwidth]{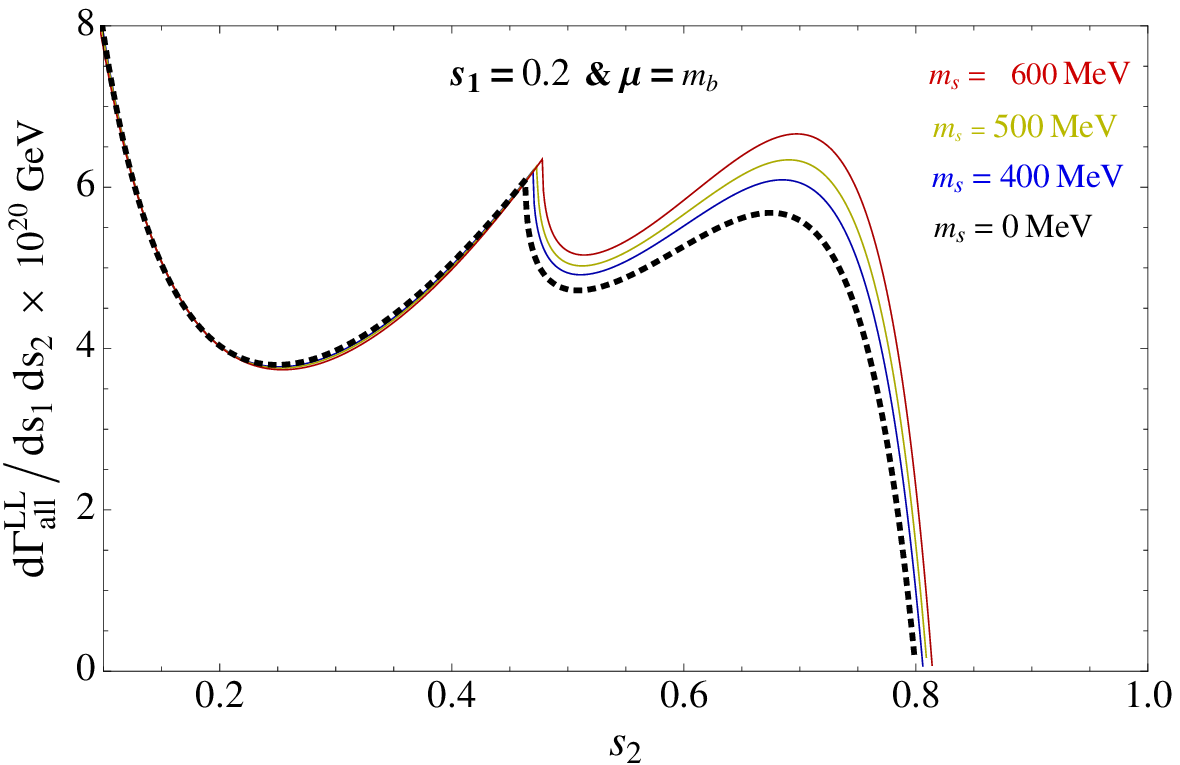}
~~~\includegraphics[width=0.42\textwidth]{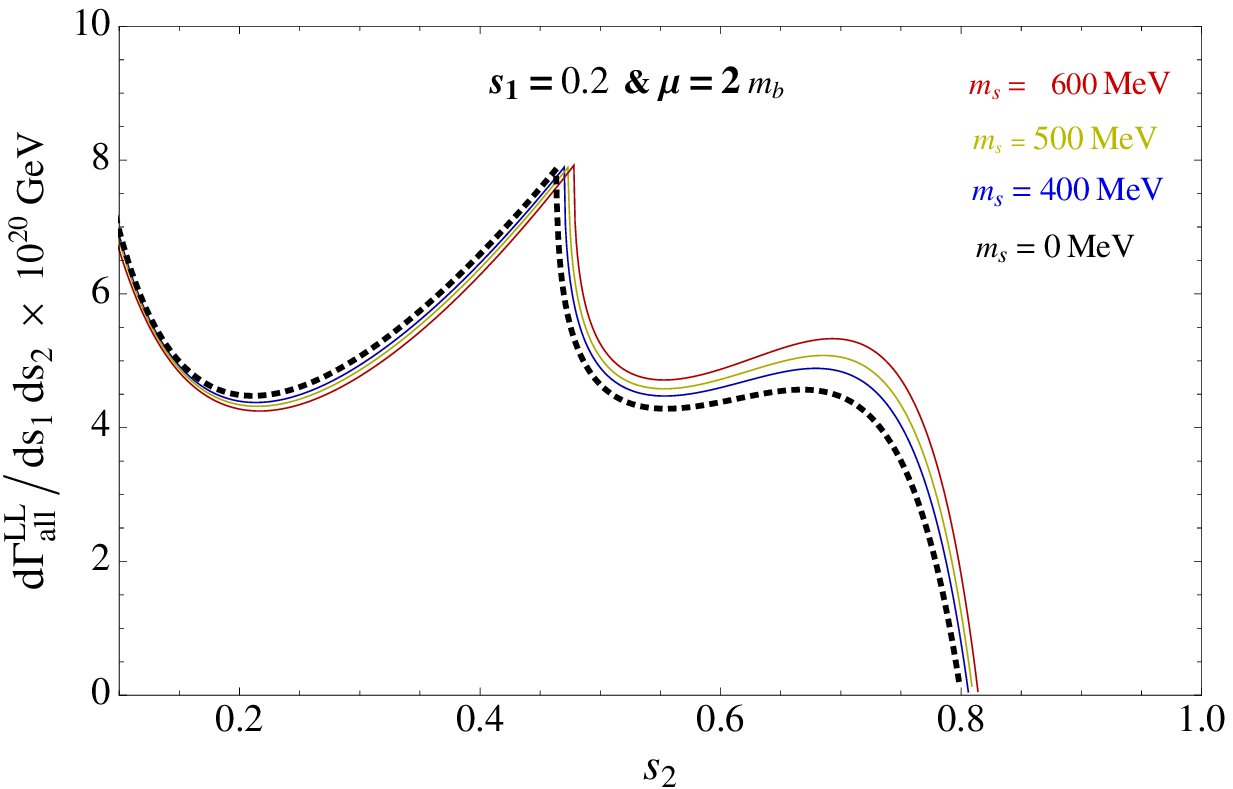}
\caption{Leading-order spectrum based on all operator contributions as given in \eq{eq:tree77A} and \eq{eq:treeRem}, as a function of $s_2$ for $s_1$ fixed at $0.2$ and $\mu \in [m_{b}/2,2 m_b]$. The dotted (lowermost), blue, yellow and red (uppermost) lines in these frames describe the results when putting $m_s=0$, $m_s=400$ MeV, $m_s=500$ MeV and $m_s=600$ MeV, respectively.}      
\label{fig:DoubleWidthLLall}
\end{center} 
\end{figure}
%


The individual ${\tilde r}_i$ quantities read
\bea
\nn {\tilde r}_a&=&
8 (s_1^2 s_2 + s_1^4 s_2 + s_1 s_2^2 - 12 s_1^2 s_2^2 + 14 s_1^3 s_2^2 - 
   7 s_1^4 s_2^2 + 14 s_1^2 s_2^3 - 24 s_1^3 s_2^3 + 12 s_1^4 s_2^3 + 
   s_1 s_2^4 - 7 s_1^2 s_2^4 + 12 s_1^3 s_2^4 \\  \nn
    &-& 6 s_1^4 s_2^4 - 3 s_1^2 x_4 + 
   2 s_1^3 x_4 - s_1^4 x_4 - 2 s_1 s_2 x_4 + 23 s_1^2 s_2 x_4 - 38 s_1^3 s_2 x_4 + 
   13 s_1^4 s_2 x_4 - 3 s_2^2 x_4 + 23 s_1 s_2^2 x_4 \\  \nn&-& 50 s_1^2 s_2^2 x_4 + 
   68 s_1^3 s_2^2 x_4 - 24 s_1^4 s_2^2 x_4 + 2 s_2^3 x_4 - 38 s_1 s_2^3 x_4 + 
   68 s_1^2 s_2^3 x_4 - 52 s_1^3 s_2^3 x_4 + 12 s_1^4 s_2^3 x_4 - s_2^4 x_4 \\  \nn&+& 
   13 s_1 s_2^4 x_4 - 24 s_1^2 s_2^4 x_4 + 12 s_1^3 s_2^4 x_4 + 5 s_1 x_4^2 - 
   7 s_1^2 x_4^2 + 16 s_1^3 x_4^2 - 6 s_1^4 x_4^2 + 5 s_2 x_4^2 - 
   56 s_1 s_2 x_4^2 + 82 s_1^2 s_2 x_4^2 \\  \nn&-& 48 s_1^3 s_2 x_4^2 + 
   13 s_1^4 s_2 x_4^2 - 7 s_2^2 x_4^2 + 82 s_1 s_2^2 x_4^2 - 
   152 s_1^2 s_2^2 x_4^2 + 68 s_1^3 s_2^2 x_4^2 - 7 s_1^4 s_2^2 x_4^2 + 
   16 s_2^3 x_4^2 - 48 s_1 s_2^3 x_4^2 \\  \nn   \nn&+& 68 s_1^2 s_2^3 x_4^2 - 
   24 s_1^3 s_2^3 x_4^2 - 6 s_2^4 x_4^2 + 13 s_1 s_2^4 x_4^2 - 
   7 s_1^2 s_2^4 x_4^2 - 4 x_4^3 + 19 s_1 x_4^3 - 42 s_1^2 x_4^3 + 
   16 s_1^3 x_4^3 - s_1^4 x_4^3 + 19 s_2 x_4^3 \\  \nn&-& 48 s_1 s_2 x_4^3 + 
   82 s_1^2 s_2 x_4^3 - 38 s_1^3 s_2 x_4^3 + s_1^4 s_2 x_4^3 - 42 s_2^2 x_4^3 + 
   82 s_1 s_2^2 x_4^3 - 50 s_1^2 s_2^2 x_4^3 + 14 s_1^3 s_2^2 x_4^3 + 
   16 s_2^3 x_4^3 \\   \nn &-& 38 s_1 s_2^3 x_4^3 + 14 s_1^2 s_2^3 x_4^3 - s_2^4 x_4^3 + 
   s_1 s_2^4 x_4^3 - 6 x_4^4 + 19 s_1 x_4^4 - 7 s_1^2 x_4^4 + 2 s_1^3 x_4^4 + 
   19 s_2 x_4^4 - 56 s_1 s_2 x_4^4 + 23 s_1^2 s_2 x_4^4 \\  \nn &-& 7 s_2^2 x_4^4 + 
   23 s_1 s_2^2 x_4^4 - 12 s_1^2 s_2^2 x_4^4 + 2 s_2^3 x_4^4 - 4 x_4^5 + 
   5 s_1 x_4^5 - 3 s_1^2 x_4^5 + 5 s_2 x_4^5 - 2 s_1 s_2 x_4^5 + 
   s_1^2 s_2 x_4^5 - 3 s_2^2 x_4^5 + s_1 s_2^2 x_4^5) \, , \\  
{\tilde r}_b&=&-32 (1 - s_1) (1 - s_2) (s_1-x_4 ) (s_2-x_4) (s_1 s_2-x_4 )(1 +
x_4 - s_1 - s_2) \, , \\
\nn {\tilde r}_c&=& {\tilde r}_a \, ,
\label{eq:r0Tilde}
\eea

   \bea
   \nn {\tilde r}_1&=& -16 s_1^2 x_4^6-16 s_2^2 x_4^6-16 s_1 x_4^6+48 s_1 s_2 x_4^6-16 s_2
   x_4^6+16 x_4^6+16 s_1 s_2^3 x_4^5+144 s_1^2 x_4^5-16 s_1^2 s_2^2
   x_4^5-112 s_1 s_2^2 x_4^5  \\ \nn
   &+& 144 s_2^2 x_4^5-208 s_1 x_4^5+16 s_1^3 s_2
   x_4^5-112 s_1^2 s_2 x_4^5+224 s_1 s_2 x_4^5-208 s_2 x_4^5+112
   x_4^5-144 s_1^3 x_4^4-32 s_1^2 s_2^3 x_4^4+128 s_1 s_2^3 x_4^4  \\ \nn
   &-& 144
   s_2^3 x_4^4+448 s_1^2 x_4^4-32 s_1^3 s_2^2 x_4^4+576 s_1^2 s_2^2
   x_4^4-928 s_1 s_2^2 x_4^4+448 s_2^2 x_4^4-448 s_1 x_4^4+128 s_1^3 s_2
   x_4^4-928 s_1^2 s_2 x_4^4    \\ \nn
   &+& 1264 s_1 s_2 x_4^4-448 s_2 x_4^4+112
   x_4^4+64 s_1^4 x_4^3+16 s_1^2 s_2^4 x_4^3-80 s_1 s_2^4 x_4^3+64 s_2^4
   x_4^3-352 s_1^3 x_4^3+48 s_1^3 s_2^3 x_4^3-544 s_1^2 s_2^3 x_4^3   \\ \nn
   &+& 880
   s_1 s_2^3 x_4^3-352 s_2^3 x_4^3+448 s_1^2 x_4^3+16 s_1^4 s_2^2
   x_4^3-544 s_1^3 s_2^2 x_4^3+1744 s_1^2 s_2^2 x_4^3-1760 s_1 s_2^2
   x_4^3+448 s_2^2 x_4^3-208 s_1 x_4^3   \\ \nn
   &-& 80 s_1^4 s_2 x_4^3+880 s_1^3 s_2
   x_4^3-1760 s_1^2 s_2 x_4^3+1264 s_1 s_2 x_4^3-208 s_2 x_4^3+16
   x_4^3+64 s_1^4 x_4^2-16 s_1^3 s_2^4 x_4^2+176 s_1^2 s_2^4 x_4^2    \\ \nn
   &-& 224
   s_1 s_2^4 x_4^2+64 s_2^4 x_4^2-144 s_1^3 x_4^2-16 s_1^4 s_2^3
   x_4^2+464 s_1^3 s_2^3 x_4^2-1152 s_1^2 s_2^3 x_4^2+880 s_1 s_2^3
   x_4^2-144 s_2^3 x_4^2+144 s_1^2 x_4^2    \\ \nn
   &+& 176 s_1^4 s_2^2 x_4^2-1152 s_1^3
   s_2^2 x_4^2+1744 s_1^2 s_2^2 x_4^2-928 s_1 s_2^2 x_4^2+144 s_2^2
   x_4^2-16 s_1 x_4^2-224 s_1^4 s_2 x_4^2+880 s_1^3 s_2 x_4^2-928 s_1^2
   s_2 x_4^2     \\ \nn
   &+& 224 s_1 s_2 x_4^2-16 s_2 x_4^2-96 s_1^3 s_2^4 x_4+176 s_1^2
   s_2^4 x_4-80 s_1 s_2^4 x_4-96 s_1^4 s_2^3 x_4+464 s_1^3 s_2^3 x_4-544
   s_1^2 s_2^3 x_4+128 s_1 s_2^3 x_4      \\ \nn
   &-& 16 s_1^2 x_4+176 s_1^4 s_2^2 x_4-544
   s_1^3 s_2^2 x_4+576 s_1^2 s_2^2 x_4-112 s_1 s_2^2 x_4-16 s_2^2 x_4-80
   s_1^4 s_2 x_4+128 s_1^3 s_2 x_4-112 s_1^2 s_2 x_4   \\ \nn
   &+&  48 s_1 s_2 x_4-16
   s_1^3 s_2^4+16 s_1^2 s_2^4-16 s_1^4 s_2^3+48 s_1^3 s_2^3-32 s_1^2
   s_2^3+16 s_1 s_2^3+16 s_1^4 s_2^2-32 s_1^3 s_2^2-16 s_1^2 s_2^2+16
   s_1^3 s_2 \, ,  \\ \nn
      \label{eq:r1Tilde} 
\eea
\be
 \nn {\tilde r}_2 = - {\tilde r}_0\, \log\left(  s_1 s_2 - x_4    \right) \,
 ,~  {\tilde r}_3 = - {\tilde r}_0\, \log\left( 1-  s_1 - s_2  + x_4
 \right)\, ,
\label{eq:r23Tilde} 
\ee
\be
 \nn {\tilde r}_0 =  {\tilde r}_a+x_4 \left( {\tilde r}_b+ {\tilde r}_c\right
 ) \, .
\label{eq:r23Tilde} 
\ee
The leading-order spectrum in $d=4$ dimensions is simply understood to be
obtained from \eq{eq:tree77A} and \eq{eq:tree77B} by setting $\epsilon$ to
zero. In the limit $m_s\rightarrow0$, \eq{eq:tree77A} correctly reproduces the
tree-level result given in Eq.~(2.3) of
Ref.~\cite{Asatrian:2014mwa}, which was derived in this limit.

For completeness, we have also improved the lowest order results for the double
differential decay width based on the remaining operators
${\cal O}_1$ and ${\cal O}_2$ by working out the full $m_s$ dependence. For this
piece we obtain\footnote{Note that in Eq.~(2.10) of
  Ref.~\cite{Asatrian:2014mwa} there was a sign mistake which is corrected
  in \eq{eq:treeRem} of the present work.}
\bea
 \frac{d\Gamma_{\rm Remaining}^{(0)}}{ds_1 \, ds_2}  &&= \frac{\alpha^2 \, m_b^5 
 \, G_F^2 \,
  |V_{tb} V_{ts}^*|^2}{1024 \, \pi^5} \, \times \left\{      4 Q_u^4 \left(   C_2(\mu)+\frac{4}{3} C_1(\mu)  \right)^2 \,  \frac{\left(    s_1+s_2  - (4-s_1-s_2)  x_4 \right)}{\left(  1-s_1-s_2+x_4 \right)^2}  \,  \tilde{h}_{(2,1)}  \right.  \nn \\ 
 && -  \left.  16 Q_u^2 Q_d \left(   C_2(\mu)+\frac{4}{3} C_1(\mu)  \right)\,
   C_{7,eff}(\mu) \, \frac{1}{ (1-s_1) (1-s_2) (s_1-x_4) (s_2-x_4)}  \,
   \tilde{h}_{(2,1,7)}           \right\}  \, . \nn \\[1mm]
\label{eq:treeRem}
\eea
The $\tilde{h}$ functions appearing in \eq{eq:treeRem} read
\be
\tilde{h}_{(2,1)} = \left|   1-s_1-s_2+x_4-4 \hat{m}_c^2 \arcsin^2(\tilde{z})     \right|^2\, ,
\ee
\bea
\nn \tilde{h}_{(2,1,7)} &=&    \left( s_1 s_2 x_4^3  -2 s_1 x_4^3    -2 s_2 x_4^3+2 s_1^2 x_4^2-s_1 s_2^2 x_4^2+2
   s_2^2 x_4^2-6 s_1 x_4^2-s_1^2 s_2 x_4^2+7 s_1 s_2 x_4^2-6 s_2 x_4^2+2
   s_1^2 x_4  \right. \\
   &+& \left. s_1^2 s_2^2 x_4-4 s_1 s_2^2 x_4+2 s_2^2 x_4-2 s_1 x_4-4
   s_1^2 s_2 x_4+7 s_1 s_2 x_4-2 s_2 x_4+s_1^2 s_2^2-s_1 s_2^2-s_1^2
   s_2+s_1 s_2 \right. \\ 
   &+& \left. 3 x_4^3+3 x_4^2  \right)\, \left( 1-s_1-s_2 + x_4 -4 \hat{m}_c^2 {\rm Re} \left[ \arcsin^2(\tilde{z}) \right]     \right)  \nn \, ,
\eea
where $\hat{m}_c=m_c/m_b$.
The argument of the $\arcsin$ function reads
$\tilde{z}=\sqrt{(1-s_1-s_2+x_4)/(4 \hat{m}_c^2)}$, 
where $\hat{m}_c^2$ is tacitly understood to have a small negative imaginary part.
In Fig.~\ref{fig:DoubleWidthLLall} we present the  leading-order spectrum based on
all operators (see \eq{eq:tree77A} and \eq{eq:treeRem})
as a function of $s_2$ for $s_1$ fixed at $0.2$ and $\mu \in [m_{b}/2,2
m_b]$. The dotted (lowermost),  blue, yellow and red (uppermost) lines in
these frames describe the results when putting $m_s=0$, $m_s=400$ MeV,
$m_s=500$ MeV and $m_s=600$ MeV, respectively.  
The numerical values of the input parameters and of the Wilson coefficients
are listed in Table \ref{tab12:wilson_input}.
We see that for $\mu=m_b/2$ the $({\cal O}_7,{\cal O}_7)$ contribution
is by far the dominant one.
This can be easily understood from  \eq{eq:treeRem}. For the renormalization
scale $\mu=m_b/2$ the combination $\left(C_2(\mu)+\frac{4}{3}
  C_1(\mu)\right)$ tends to zero since $C_2(m_b/2) \approx
-\frac{4}{3}C_1(m_b/2)$ and thus the $({\cal O}_{7},{\cal O}_{7})$
contribution dominates at this scale.

\begin{table}[htbp]
\begin{minipage}{3in}
\centering 
  \begin{tabular}{@{}|c|c|c|c@{}|}
 \hline
 Parameter & Value \\   \hline \hline

$BR_{sl}^{exp}$& $0.1049$   \\ \hline

$m_{c}/m_{b}$ & $0.29$   \\ \hline

$m_{b}$& $4.8$~GeV   \\ \hline

$m_{t}$&$175$~GeV    \\ \hline

$m_{W}$&$80.4$~GeV    \\ \hline

$m_{Z}$&$91.19$~GeV   \\ \hline

$G_{F}$&$1.16637\times10^{-5}$~\text{GeV}$^{-2}$  \\ \hline

$V_{cb} $&$0.04$    \\  \hline

$V_{tb} V_{ts}^* $&$0.04$    \\  \hline

${\alpha_{\rm (em)}}^{-1}$ & $137$    \\ \hline

${\alpha_{s}(m_{Z})}$&$0.119$      \\ \hline \hline
\end{tabular}
 \end{minipage} ~
\begin{minipage}{3in}
  \vspace{0.5cm}
  \begin{tabular}{|c|c|c|c|c|c|}
\hline
 & $C_{1}^{0}(\mu)$ ~&~ $C_{2}^{0}(\mu)$ ~&  $C_{7,eff}^{0}(\mu)$ ~&~ $C_{7,eff}^{1}(\mu)$ &  $\alpha_s(\mu)$ 
\\   \hline \hline

$\mu= {m_W}$  &$0$  &$1$ &$-0.1957$  &$-2.3835$    &$0.1213$    \\ \hline 

 $\mu=2 \, {m_{b}}$ &$-0.3352$  &$1.0116$ & $-0.2796$  &$-0.1788$   &$0.1818$    \\ \hline 

 $\mu={m_{b}}$  &$-0.4976$ &$1.0245$ & $-0.3142$ &  $0.4728$  &$0.2175$    \\ \hline

 $\mu={m_{b}}/2$  &$-0.7117$  &$1.0478$ &  $-0.3556$ & $1.0794$    &$0.2714$         \\ \hline \hline
\end{tabular}  
    \end{minipage}
\caption{ {\bf Left:} Input parameters used in this paper. {\bf Right:} Relevant Wilson coefficients and $\alpha_s(\mu)$ at
   different values of the renormalization scale $\mu$.}
\label{tab12:wilson_input}
\end{table}   
\vspace{1mm}

\section{Improved ${\cal O}(\alpha_{s})$ results for the double differential spectrum
  $d\Gamma/(ds_1 ds_2)$}
\label{sec:NLL}
\subsection{Virtual corrections}
\label{subsec:Virtual}
\begin{figure}[h]
\begin{center}
\includegraphics[width=0.9\textwidth]{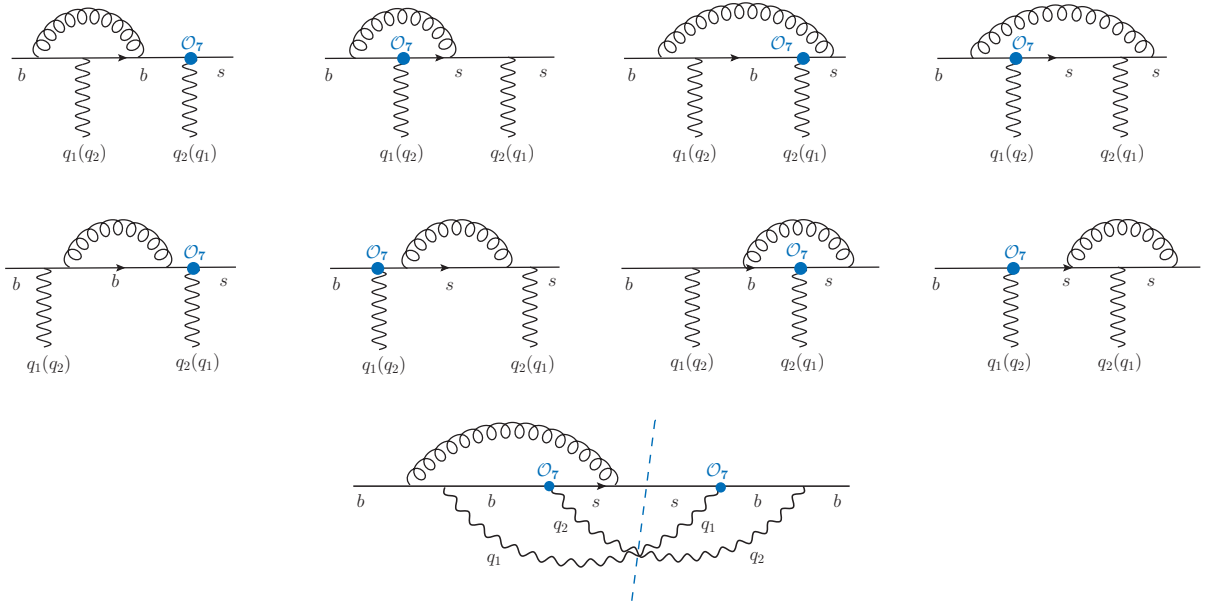}
\caption{On the first two lines the one-loop Feynman diagrams for 
$b \to s \gamma \gamma$ associated with ${\cal O}_7$ are
shown at the amplitude level. Diagrams with self-energy insertions on the external quark-legs
are not shown. On the last line the contribution to the decay width corresponding
to the interference of the third diagram on the first line with the first
(tree-level) diagram in Fig. \ref{fig:TreeDiags} with $q_1\leftrightarrow q_2$ is shown.}
\label{fig:3PcutDiags}
\end{center}
\end{figure}

We now turn to the calculation of the virtual QCD corrections, i.e. to
the contributions of order $\alpha_s$ with three particles in the
final state. The diagrams defining the (unrenormalized) 
virtual corrections at the amplitude level are shown in the first two lines of
Fig. \ref{fig:3PcutDiags}. 
As the diagrams with a self-energy insertion on the external $b$- and
$s$-quark legs are taken into account in the renormalization process,
these diagrams are not shown in Fig. \ref{fig:3PcutDiags}.
In order to get the (unrenormalized) virtual corrections 
$d\Gamma_{77}^{\rm bare}/(ds_1 ds_2)$ of order $\alpha_s$, 
we have to work out the interference of the diagrams 
in Fig. \ref{fig:3PcutDiags} with the leading
order diagrams in Fig. \ref{fig:TreeDiags}.

From the technical point of view we use two different methods to perform
the calculations.
In the first method we use the Laporta Algorithm \cite{Laporta:2001dd}  
(see also \cite{Tkachov:1981wb,Chetyrkin:1981qh})
to identify the needed Master Integrals, followed by applying the differential equation method to
solve them. As we used these
techniques also in \cite{Asatrian:2011ta}, we refer
to section 7 of that paper which contains the technical details
and the corresponding references.
In the second method, the one-loop amplitudes are reduced to tensor integrals
and subsequently decomposed into their Lorentz-covariant structure by means of
the {\it Mathematica} package FEYNCALC \cite{Mertig:1990an,Shtabovenko:2016sxi}.
For the numerical evaluation of the
tensor-coefficient functions we employed the LOOPTOOLS library
\cite{Hahn:1998yk,vanOldenborgh:1989wn}. For some checks, we also used the
SecDec-3.0 package \cite{Borowka:2015mxa}.
We note that the two methods give the same result, providing us with firm check of our results.

In order to renormalize the calculated bare $O(\alpha_s)$ virtual corrections,
one needs to add counterterm contributions which, in our case, can be divided into two parts as
\be
\frac{d\Gamma_{77}^{\rm ct}}{ds_1 ds_2}=
\frac{d\Gamma_{77}^{{\rm ct},(A)}}{ds_1 ds_2}+
\frac{d\Gamma_{77}^{{\rm ct},(B)}}{ds_1 ds_2} \, . 
\ee
\noindent
Part (A) involves the Lehmann, Symanzik, Zimmermann (LSZ) factors 
$\sqrt{Z_{2b}^{\rm OS}}$ and $\sqrt{Z_{2s}^{\rm OS}}$ for the $b$- and
$s$-quark fields, as well as the
self-renormalization constant $Z_{77}^{\MS}$ of the operator ${\cal
  O}_7$, as well as $Z_{m_b}^{\MS}$ and $Z_{m_s}^{\MS}$
renormalizing the factors $\bar{m}_b(\mu)$ and $\bar{m}_s(\mu)$ present in the
operator ${\cal O}_7$. Defining $\delta Z_i= Z_i-1$, we get for part (A)
\be
\frac{d\Gamma_{77}^{{\rm ct},(A)}}{ds_1 ds_2} = 
\left[ \delta Z_{2b}^{\rm OS} + \delta Z_{2s}^{\rm OS} + 2 \, \delta
  Z_{m}^{\MS} +
  2 \, \delta Z_{77}^{\MS} \right] \, \frac{d\Gamma_{77}^{(0,d)}}{ds_1 ds_2} \, .
\ee
The simple structure of this result is related to the fact that the 
$\MS$ renormalization constants of the bottom and the strange quark mass
are identical, i.e., $Z_{m_b}^{\MS}=Z_{m_s}^{\MS} \equiv Z_{m}^{\MS}$.

\begin{center}
\begin{figure}[h]
\centering{
\includegraphics[width=0.55\textwidth]{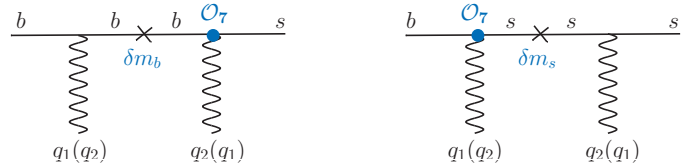}
\caption{Counterterm amplitudes involving $\delta m_{s,b}$ insertions in the internal quark lines.}      
\label{fig:CTB}
}
\end{figure}
\end{center}

The counterterms defining part (B) are due to the insertion 
of $-i \delta m_b \bar b b$ and $-i \delta m_s \bar s s$  
in the internal $b$ and $s$-quark lines in the leading order diagrams as
indicated in Fig. \ref{fig:CTB}, where 
\[
\delta m_b = (Z_{m_b}^{\rm OS}-1) \, m_b, \,~ \delta m_s = (Z_{m_s}^{\rm
  OS}-1) \, m_s \, .
\]
More precisely, part (B) consists of the interference of the diagrams in
Fig.~\ref{fig:CTB} with the leading order diagrams in 
Fig.~\ref{fig:TreeDiags}. The various $Z$-factors are listed for completeness
in Appendix \ref{subsec:Renfactors}.

By adding $d\Gamma_{77}^{\rm bare}/(ds_1 ds_2)$ and 
$d \Gamma_{77}^{{\rm ct}}/(ds_1 ds_2)$,
we get the result for the renormalized virtual corrections to the
spectrum, $d\Gamma_{77}^{(1),virt}/(ds_1 \, ds_2)$.

\be
\frac{d\Gamma_{77}^{(1),virt}}{ds_1 \, ds_2} = 
\frac{d\Gamma_{77}^{\rm bare}}{ds_1 ds_2}+\frac{d \Gamma_{77}^{{\rm ct}}}{ds_1
  ds_2} \, .
\ee

%

\subsection{Bremsstrahlung corrections}
\label{subsec:Bremsstrahlung}

\begin{figure}[h]
\begin{center}
\includegraphics[width=0.45\textwidth]{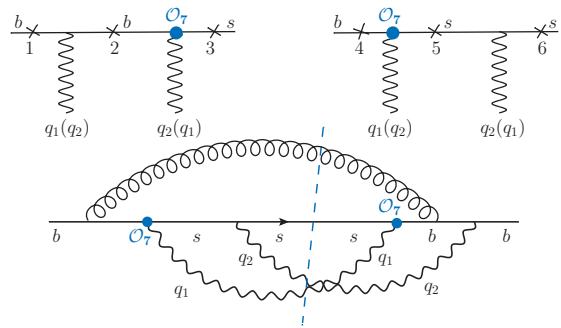}
\caption{On the first line the diagrams defining the
${\cal O}_7$ contribution to $b \to s g \gamma \gamma$ are shown
at the amplitude level. The crosses in the graphs stand
for the possible emission places of the gluon.
On the second line the contribution to the decay width corresponding
to the interference of  diagram 4 with diagram 2 ($q_1\leftrightarrow q_2$) is illustrated. This sample
interference diagram gives rise to
$\log\left( m_{s}/m_{b}\right)$ terms due to collinear configurations of one of the photons with the $s$-quark.}
\label{fig:4PcutDiags}
\end{center}
\end{figure}

We now turn to the calculation of the bremsstrahlung QCD corrections, i.e. to
the contributions of order $\alpha_s$ with four particles in the
final state.
The corresponding diagrams at the
amplitude level are shown on the first line in Fig. \ref{fig:4PcutDiags}. 
We use again two methods to calculate the bremsstrahlung corrections.
In the first one, we  use  the Laporta Algorithm \cite{Laporta:2001dd}  
to identify the Master Integrals, which are then solved by applying the
differential equation method.  
As in \cite{Asatrian:2013raa} we worked out in a first step the triple
differential spectrum $d\Gamma_{77}^{(1),brems}/(ds_1 ds_2 ds_3)$,  $s_3=(p_s+p_g)^2/m_b^2$,
obtaining a fully analytic result which however is very lengthy. 
To get the double differential spectrum $d\Gamma_{77}^{(1),brems}/{ds_1 ds_2}$, we integrated over $s_3$, 
which runs in the interval $[m_s^2/m_b^2,s_1 \, s_2]$. In some terms this integration was
done numerically.
In the second method we perform a slicing of the phase-space: We introduce a
small gluon energy cut-off $\omega_0$ and divide the real emission
contribution into a {\it soft} and a {\it hard} part. The soft part, which
contains the infrared (IR) singularity, comes from the phase space region where the
gluon energy is below $\omega_0$. As in the case of the virtual diagrams, the IR
singularities are regularized dimensionally. By taking advantage of the soft
gluon approximation for the amplitude of the bremsstrahlung process, it is
possible to perform the integral with respect to the gluon momentum
analytically; the process-independent result, which was derived in 
\cite{Hooft:1978xw} (see also \cite{Denner:1991kt}), 
depends only on the momenta of the external particles in
the corresponding Born process. We explicitly checked that the IR divergencies
cancel out once the soft part is combined with the virtual corrections.
In order to obtain the hard part contribution to the double differential decay
width, a three-dimensional (finite) integral is involved; we performed this
integration numerically, by making use of the CUBA-library \cite{Hahn:2004fe}.    
Again, the two methods lead to the same result.

%

\subsection{Improved final result for the decay width at ${\cal O}(\alpha_{s})$}
\label{subsec:Finalres}
The complete order $\alpha_s$ correction to the
double differential decay width
$d\Gamma_{77}/(ds_1 \, ds_2)$ is obtained by
adding the renormalized virtual corrections from section
\ref{subsec:Virtual} and the bremsstrahlung corrections discussed in
section \ref{subsec:Bremsstrahlung}:
\be
\frac{d\Gamma_{77}^{1}}{ds_1 ds_2} = 
\frac{d\Gamma_{77}^{(1),virt}}{ds_1 ds_2} +
\frac{d\Gamma_{77}^{(1),brems}}{ds_1 ds_2} \, .
\label{alphas77}
\ee

\section{Numerical illustrations}\label{sec:numerics}
The NLL prediction for the double differential decay width reads
\be
\frac{d\Gamma_{77}}{ds_1 ds_2} = 
\frac{d\Gamma_{77}^{(0,4)}}{ds_1 ds_2} + \frac{d\Gamma_{77}^{(1)}}{ds_1 ds_2}
\, .
\label{total77}
\ee 

To illustrate our results, we first rewrite the $\overline{\mbox{MS}}$ masses
$\bar{m}_b(\mu)$, $\bar{m}_s(\mu)$ in \eq{total77} in terms of the pole masses
$m_b$,  $m_s$, using the one-loop relations

\be
\bar{m}_b(\mu) = m_b \, \left[ 1 - \frac{\alpha_s(\mu)}{4 \pi} \, \left( 8 \log
  \frac{\mu}{m_b} + \frac{16}{3} \right)
\right] \, ,
\ee

\be
\bar{m}_s(\mu) = m_s \, \left[ 1 - \frac{\alpha_s(\mu)}{4 \pi} \, \left( 8 \log
  \frac{\mu}{m_s} + \frac{16}{3} \right)
\right] \, .
\ee

\noindent
We then insert $C_{7,eff}(\mu)$ in the form

\be
C_{7,eff}(\mu) = C_{7,eff}^{0}(\mu) + \frac{\alpha_s(\mu)}{4\pi} \, C_{7,eff}^{1}(\mu)
\label{wilsonexpand}
\ee
and expand the resulting expression for
$d\Gamma_{77}/(ds_1 ds_2)$ w.r.t. $\alpha_s$, discarding terms of
order $\alpha_s^2$. This procedure defines the NLL result. The
corresponding LL result is obtained by discarding the order
$\alpha_s^1$ terms. The numerical values for
the input parameters and for the Wilson coefficient $C_{7,eff}(\mu)$ 
at various values for the scale $\mu$, together with the
numerical values of $\alpha_s(\mu)$, are given in Table \ref{tab12:wilson_input}.

\begin{figure}[h]
\begin{center}
\includegraphics[width=0.42\textwidth]{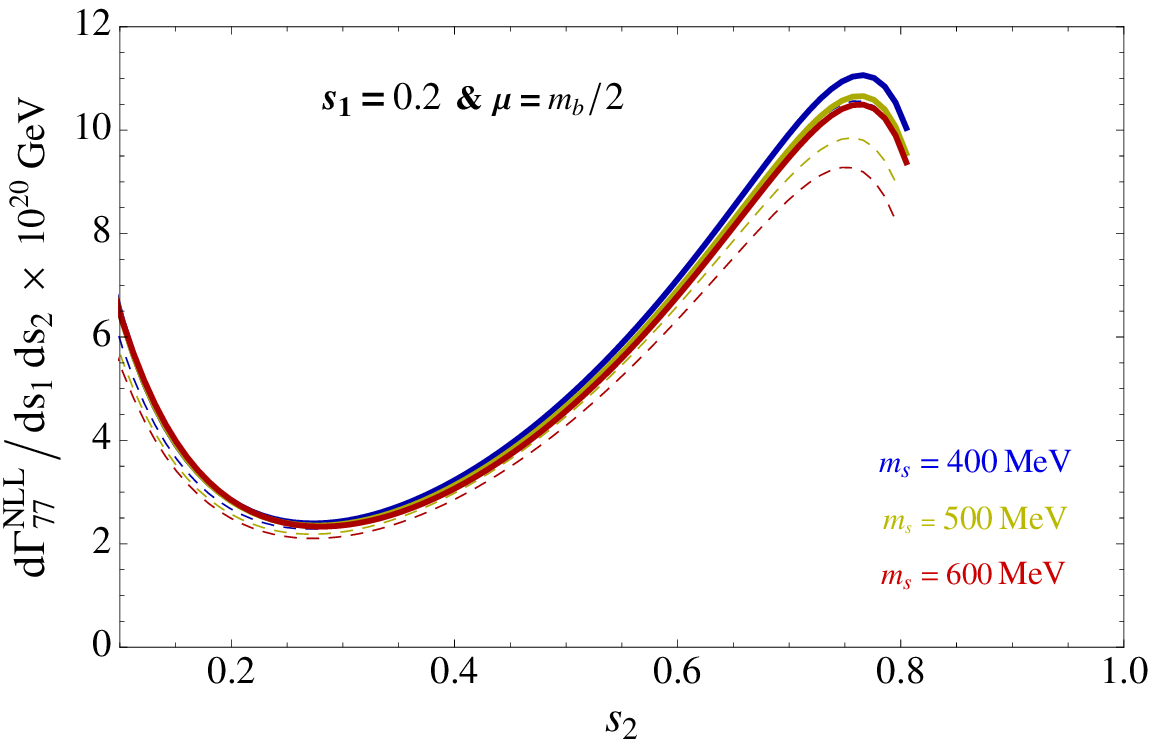}
~~~~~~~~~~~~~~~~~\includegraphics[width=0.41\textwidth]{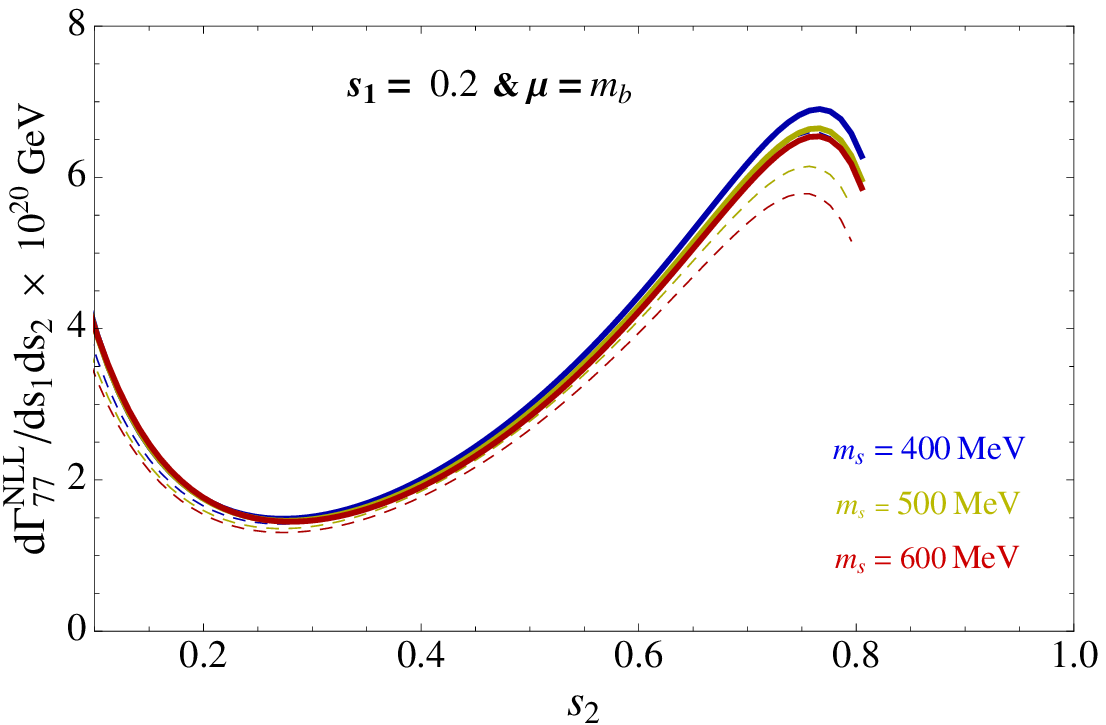} 
~~~~~~~~~~~~~~~~~\includegraphics[width=0.41\textwidth]{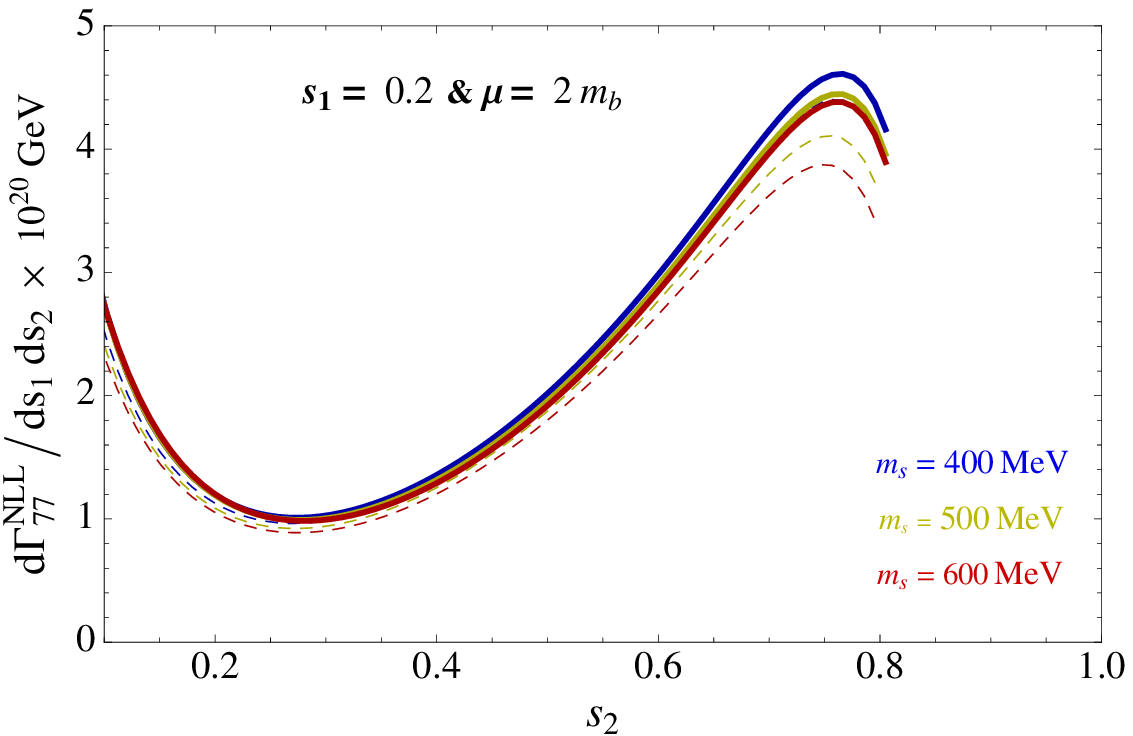} 
\caption{NLL double differential spectrum based on (${\mathcal
    O}_7$,\,${\mathcal O}_7$) contribution only, as a function of $s_2$ for
  $s_1$ fixed at $0.2$. In each of these plots, the solid curves show the
  results based on the new calculation with exact $m_s$, while the dashed
  curves are based on previous results, i.e. in the limit $m_s\to0$. See the text for details.}      
\label{fig:DoubleWidthNLLo7}
\end{center} 
\end{figure}

In Fig. \ref{fig:DoubleWidthNLLo7} we give the NLL double differential
spectrum based on the (${\mathcal O}_7$,\,${\mathcal O}_7$) contribution only, as
a function of $s_2$ for $s_1$ fixed at $0.2$. In each of these plots, the
solid curves show the results based on the present calculation with exact
$m_s$ dependence,
while the dashed curves are based on the previous approximated
result of Ref. \cite{Asatrian:2014mwa}, where only logarithmic and constant
terms in $m_s$ were kept (which we denote as ``$m_s\to 0$ results''). The
renormalization scale $\mu$ and $m_s$ are varied as explicitly displayed. A
straightforward comparison between the solid ($m_s$ exact results) and the
dashed curves ($m_s\to 0$ results) shows that finite $m_s$ effects are only
sizable near the kinematical endpoints of the spectrum. One also can see that
the exact $m_s$ results only develop a sizable $m_s$ dependence near the
kinematical endpoints of the spectrum.


\begin{figure}[h]
\begin{center}
\includegraphics[width=0.42\textwidth]{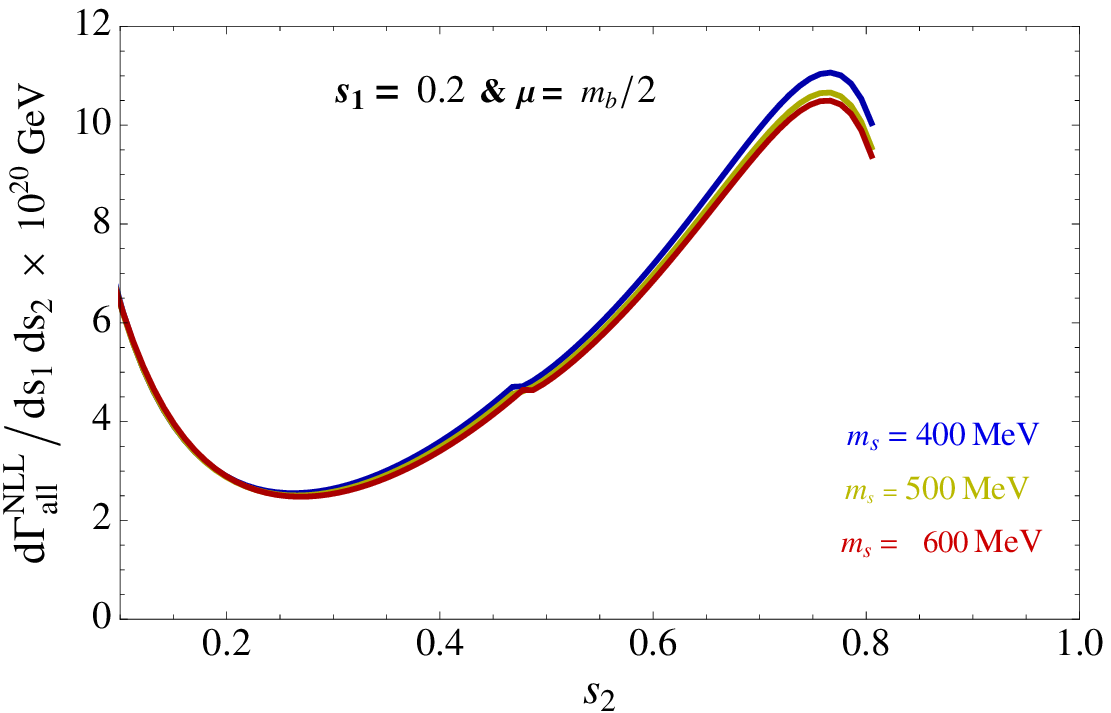}
~~~~~~~~~~~~~~~~~\includegraphics[width=0.41\textwidth]{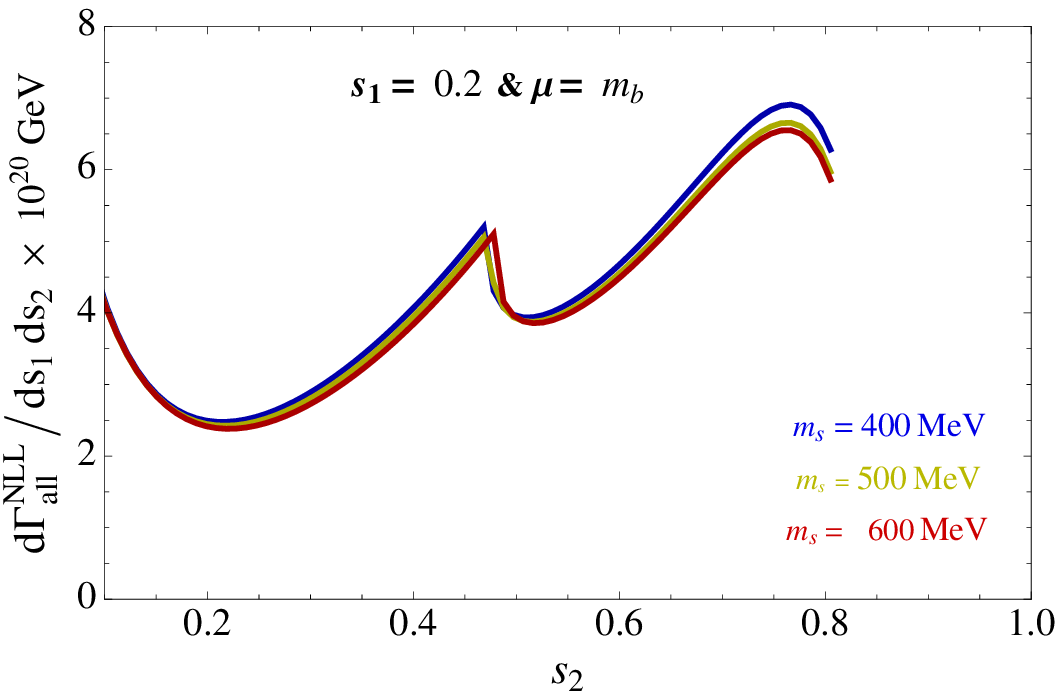} 
~~~~~~~~~~~~~~~~~\includegraphics[width=0.41\textwidth]{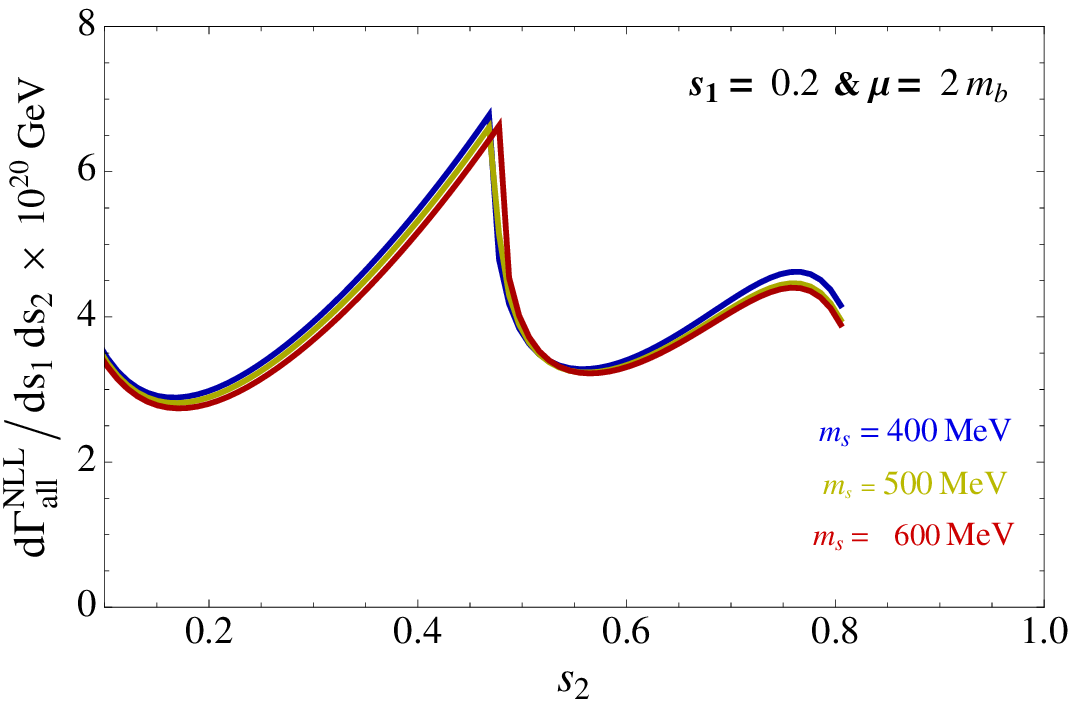} 
\caption{NLL spectrum (with exact treatment of $m_s$) based on all available operator contributions to date,
  as a function of $s_2$ for $s_1$ fixed at $0.2$. 
The blue (uppermost), yellow and red (lowermost) curves in these frames describe the results when putting $m_s=400$ MeV, $m_s=500$ MeV and $m_s=600$ MeV, respectively.} 
\label{fig:DoubleWidthNLLall}
\end{center} 
\end{figure}

In Fig. \ref{fig:DoubleWidthNLLall} we give the NLL double differential spectrum based on all
available operator contributions to date as a function of $s_2$ (for $s_1$
fixed at $0.2$) taking $\mu \in \left\{ m_{b}/2,m_b,2 m_b \right\}$ and putting 
$m_s=400$ MeV, $m_s=500$ MeV and $m_s=600$ MeV.

As the analytic expressions for the double differential decay width $d\Gamma(B \to
X_s \gamma \gamma)/(ds_1 \, ds_2)$ are very lengthy, they cannot be given in
this paper. In order to provide nevertheless the
complete information, we decided to give it in form
of an appended Fortran program, called ``doublediff.F''. The input/output information
is described in a few comment lines at the beginning of the very short main
program. The program uses the LOOPTOOLS-library
\cite{Hahn:1998yk,vanOldenborgh:1989wn} and the CUBA-library
\cite{Hahn:2004fe}. The code has been tested when using
the versions LoopTools-2.13 and Cuba-3.2, respectively.

We already mentioned that one gets a sizable $m_s$ dependence only near the kinematical endpoints 
of the spectrum. This means that if one is sufficiently away from these
endpoints, one gets more reliable predictions.  
From \eq{eq:ps2cuts} one can see that this situation can be achieved when
choosing a value for $c$ which is not too small. On the other hand, the
decay width for $\bar{B}\to X_s \gamma \gamma$ will decrease for
increasing values of $c$. It is therefore necessary to take compromising
values for $c$. Explicit calculations show that for $c=1/50$
the decay width does not develop a sizable $m_s$ dependence:
When varying  $m_s$ between $400$ and $600$ MeV, the impact on the
decay width is less than $5\%$. For larger choices of $c$, the $m_s$
sensitivity is even much smaller. 
To illustrate the $c$-dependence of the decay width, 
we use the values $c=1/50$, $c=1/25$ and $c=1/15$ in the following. 

\begin{figure}[h]
\begin{center}
\includegraphics[width=0.45\textwidth]{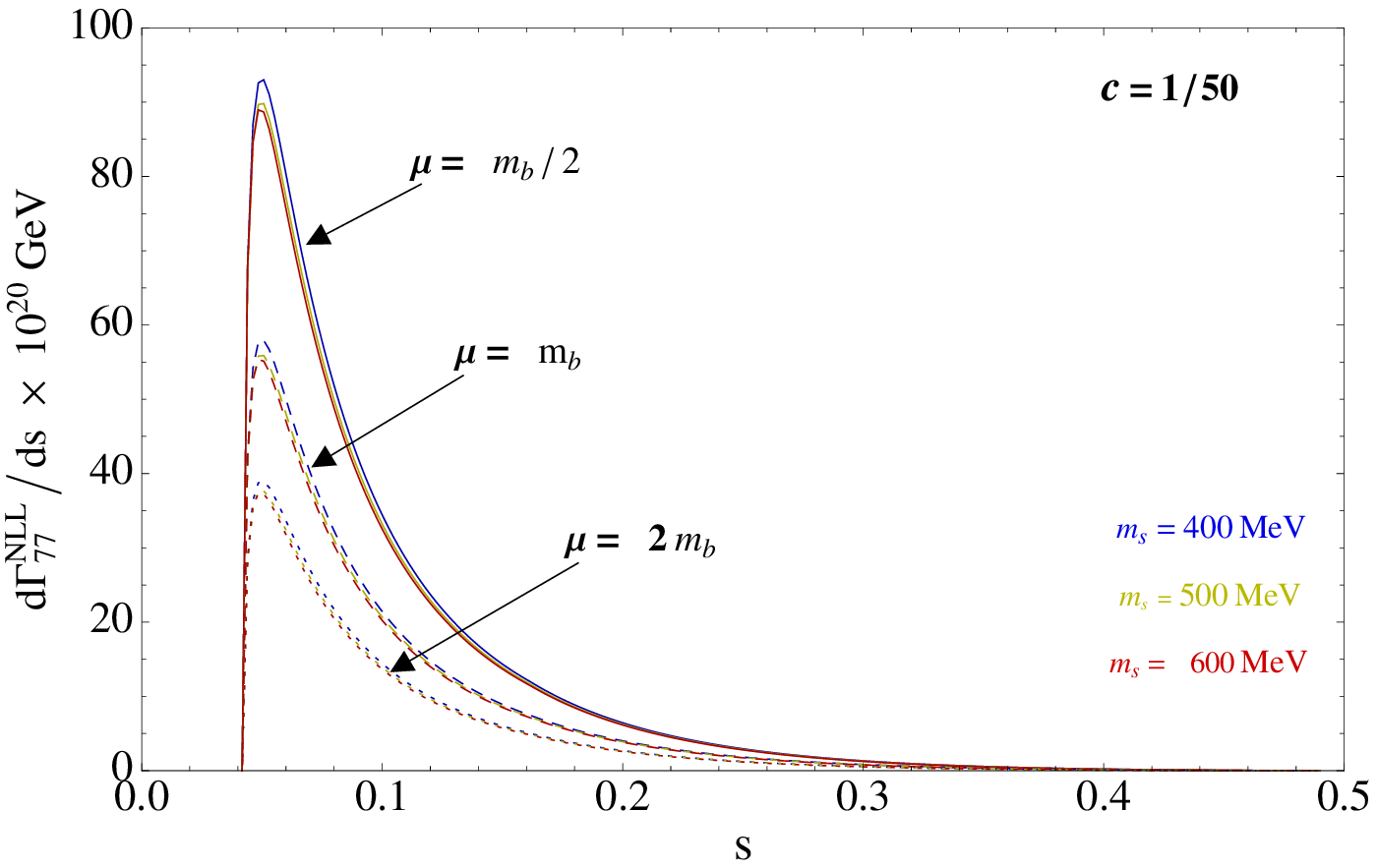}
~~~\includegraphics[width=0.45\textwidth]{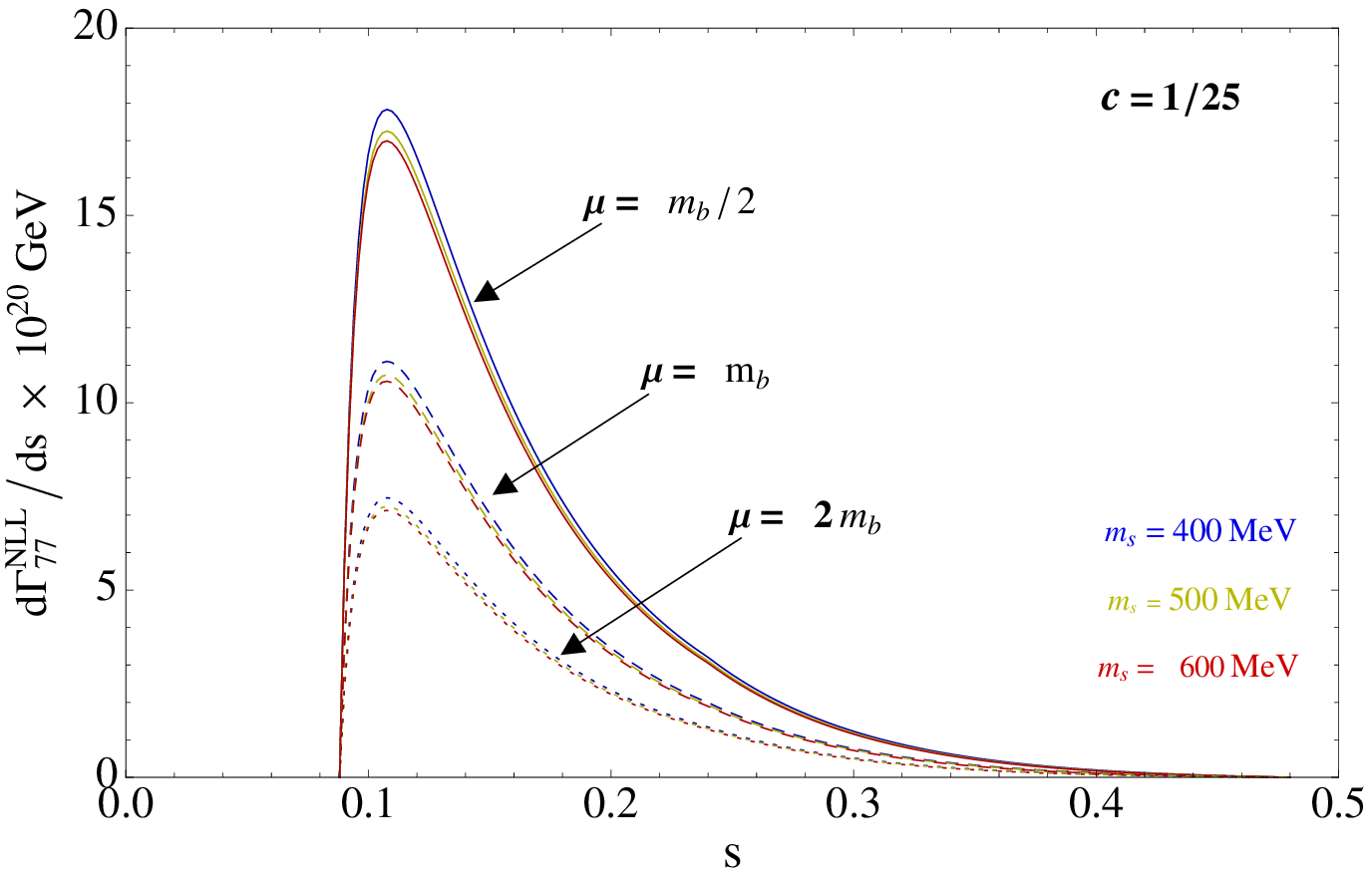}
~~~\includegraphics[width=0.45\textwidth]{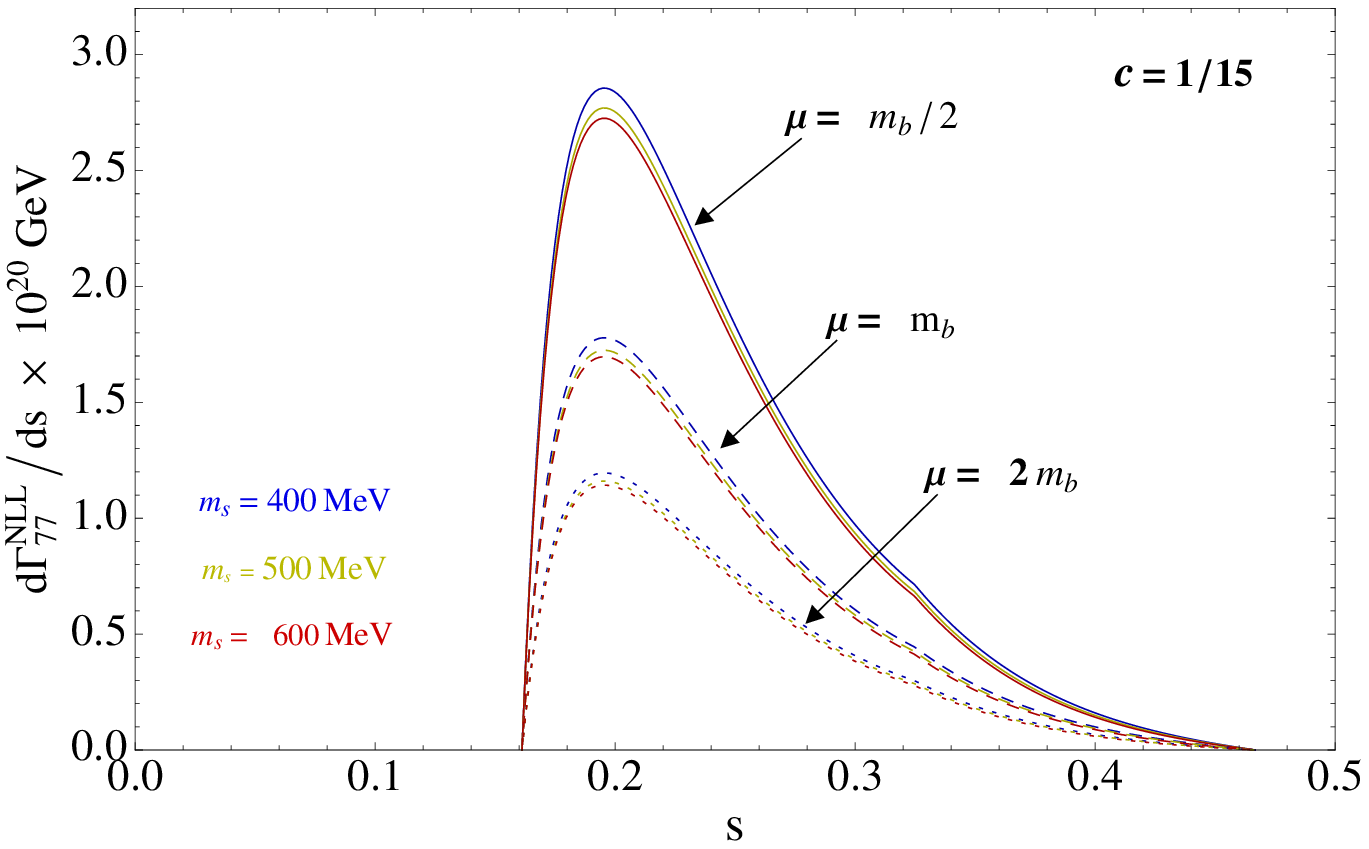}\\ 
\caption{$d\Gamma_{77}/ds$ at next-to-leading-order for different values of
  the cut-off parameter $c$.
The first frame corresponds to $c=1/50$, the second to $c=1/25$ and the third
to $c=1/15$. The blue (uppermost), yellow and red (lowermost) curves in these
plots describe the results when setting $m_s=400$ MeV, $m_s=500$ MeV and
$m_s=600$ MeV, respectively. Further, the solid (uppermost three), dashed
(middle three) and dotted (lowermost three) curves in these frames define the
results when choosing $\mu=m_b/2$, $\mu=m_b$ and $\mu=2m_b$, respectively.
See text for details. } 
\label{fig:singleWidthNLLo7}
\end{center}
\end{figure}

In the decay $B \to X_s \gamma \gamma$ two photons are emitted, characterized by
the kinematical variables $s_1$ and $s_2$. We now define the one-dimensional
physical spectrum 
\be
\frac{d\Gamma_{77}}{ds}  \qquad \qquad \mbox{where $s=\rm{min} \left\{ s_1,s_2
  \right\}$}
\, .
\ee
This observable can be constructed from the double differential spectrum
$d\Gamma_{77}/(ds_1 \, ds_2)$ in the following way: 
\be 
\frac{d\Gamma_{77}}{ds} = 2 \, \left[ \int_G \, ds_2 \,  d\Gamma_{77}/(ds_1 \,
  ds_2) \right]_{s_1 \to s} \, .
\ee
The integration interval $G$ can be specified as follows: For a given value of $s_1$
the variable $s_2$ runs over all values in the cut phase space (characterized
by \eq{eq:ps2cuts}) which satisfy the additional condition $s_2>s_1$. 
More explicitly, this  can be summarized as
\be
\frac{1}{2}(1-c-\sqrt{1-10 c+9c^2})<s_1<\frac{1}{2}(1-c) \, ; \quad \rm{max}
\left\{ s_1,c+\frac{c}{s_1-c} \right\} <s_2<1-c-s_1 \, .
\ee
In Fig. \ref{fig:singleWidthNLLo7} we show the next-to-leading order
prediction of $d\Gamma_{77}/ds$ for different values of the cut-off
paramter  $c$.  

To get the branching ratio for $\bar{B} \to X_s \gamma \gamma$ as a
function of the cut-off parameter $c$ defined in
\eq{eq:ps2cuts}, we integrate the double differential
spectrum over the corresponding range in $s_1$ and $s_2$, divide by
the semileptonic decay width and multiply with the measured
semileptonic branching ratio. 
The relevant formula for the semileptonic decay width at lowest order (which is sufficient for the purpose of this paper) reads (recalling that $\hat{m}_c=m_c/m_b$)
\be
\Gamma_{\rm sl} = \frac{m_b^5 \, G_F^2 \, |V_{cb}^2|}{192\pi^3}\, {g}(\hat{m}_c)\, ,
\ee
where the phase-space factor is defined as
\be
 g(x) = 1- 8 {x}^2  + 8 {x}^6  - x^8 -  24 \, {x}^4 \log({x})\, .
\ee

Using the input parameters in Table 
 \ref{tab12:wilson_input}, we get the branching ratios shown in Table
\ref{tab:o7o7BRs} for different values of the cut-off parameter $c$.

{  \begin{widetext}
\begin{center}
{ \renewcommand{\arraystretch}{1.35}
\begin{table}[]
\centering
\begin{tabular}{|l|c|c|c|c|c|c || c|c|c|c|c|c|  | c|c|c|c|c|c|c| }
\hline
                                      & \multicolumn{18}{c|}{  \textcolor{Blue}{ Branching ratios for $\bar{B}\to X_s\gamma \gamma$ }   }                                                                                                                                                                                                                      \\ \hline \hline
                                      & \multicolumn{6}{c||}{$  \textcolor{Blue}{c=1/50} $}                                                                                                                 & \multicolumn{6}{c||}{$   \textcolor{Blue}{c=1/25} $}                       & \multicolumn{6}{c|}{$   \textcolor{Blue}{c=1/15} $}                                                                                          \\ \hline
                                      &\multicolumn{2}{c|}{$~\mu=\,$$m_{b}/2~$}        & \multicolumn{2}{c|}{$\mu=\,$$m_{b}$}          & \multicolumn{2}{c||}{$\mu=\,$$2m_{b}$}         & \multicolumn{2}{c|}{$~\mu=\,$$m_{b}/2~$}        & \multicolumn{2}{c|}{$\mu=\,$$m_{b}$}          & \multicolumn{2}{c||}{$\mu=\,$$2m_{b}$}       & \multicolumn{2}{c|}{$\mu=\,$$m_{b}/2$}        & \multicolumn{2}{c|}{$\mu=\,$$m_{b}$}          & \multicolumn{2}{c|}{$\mu=\,$$2m_{b}$}     \\ \hline
                                      & $~{\mathcal O}_7~$    & $~{\rm all}~$         & $~{\mathcal O}_7~$    & $~{\rm all}~$         & $~{\mathcal O}_7~$    & $~{\rm all}~$   & $~{\mathcal O}_7~$    & $~{\rm all}~$         & $~{\mathcal O}_7~$    & $~{\rm all}~$         & $~{\mathcal O}_7~$    & $~{\rm all}~$      & $~{\mathcal O}_7~$    & $~{\rm all}~$         & $~{\mathcal O}_7~$    & $~{\rm all}~$         & $~{\mathcal O}_7~$    & $~{\rm all}~$         \\ \hline \hline
\multicolumn{1}{|c|}{\textcolor{Blue}{${\rm LL}$}}      &   $0.94$                     & $0.95$                     &  $0.74$       & $0.79$     &  $0.58$                     & $0.69$       & $0.28$                     & $0.29$                     & $0.22$                     & $0.25$                     & $0.17$         & $0.24$    &0.054   &0.056    &0.042  &0.049   &0.034  &0.046               \\ \hline
\multicolumn{1}{|c|}{~\textcolor{Blue}{${\rm LL_1}$}}&   1.05                     &  1.06                     & 0.82                    &  0.87     &  0.65                     & 0.76                     & 0.30                     & 0.31                     & 0.24                     & 0.27                     & 0.19      & 0.25    &0.058   &0.059    &0.045   &0.052     &0.036  &0.049                   \\ \hline
\multicolumn{1}{|c|}{~\textcolor{Blue}{${\rm LL_2}$}}& 1.11                     & 1.12                     &0.87                     & 0.92                     & 0.69                     & 0.79                     & 0.31                     & 0.32                    & 0.25                     & 0.28                     & 0.19   & 0.26     &0.059   &0.061   &0.046   &0.054    &0.037    &0.051      \\ \hline
\multicolumn{1}{|c|}{~\textcolor{Blue}{${\rm LL_3}$}}& 1.20                     & 1.20                     & 0.93                     & 0.99                     & 0.74                     & 0.85                     & 0.33                     & 0.34                    & 0.26                     & 0.29                     & 0.20     & 0.27   &0.062     &0.064    &0.048    &0.056    &0.038  &0.053                    \\  \hline \hline
\multicolumn{1}{|c|}{\textcolor{Blue}{${\rm NLL_1}$}} &  1.18         &       1.19       & 0.73     & 0.79       & 0.49              &  0.60     &  0.35     &  0.35   &  0.22   &  0.25       &   0.15  &0.21         &0.068   &0.069    &0.042   &0.050  &0.028  & 0.042    \\ \hline
\multicolumn{1}{|c|}{\textcolor{Blue}{ ${\rm NLL_2}$}} & 1.14    &   1.15      &     0.71  & 0.76      &  0.48      &  0.58    &   0.33  &    0.34   & 0.21      &    0.24 &    0.14  &  0.21      & 0.066    & 0.067   & 0.041  & 0.049   & 0.027  & 0.041     \\ \hline
\multicolumn{1}{|c|}{\textcolor{Blue}{ ${\rm NLL_3}$}} &   1.12  &      1.13     &       0.69 &     0.75     &     0.47 &     0.58  &   0.33 &  0.34    & 0.20 &  0.24     &  0.14  & 0.20      & 0.064   & 0.066  & 0.040   & 0.048   & 0.027  & 0.042  \\ \hline \hline
\end{tabular}
\caption{Branching ratios (in units of $10^{-7}$) for $\bar{B}\to X_s\gamma
  \gamma$. The left panel of the table corresponds to the results when
  choosing the kinematical cut-off parameter $c=1/50$, the middle panel is for
  $c=1/25$ and the right one for $c=1/15$. The rows labeled as ${\rm LL}$,
  ${\rm LL_1}$, ${\rm LL_2}$ and ${\rm LL_3}$ stand for the improved
  leading-order results when setting $m_s=0$, $m_s=400$ MeV, $m_s=500$ MeV and
  $m_s=600$ MeV, respectively. The
  rows labeled with ${\rm NLL_{1}}$, ${\rm NLL_{2}}$ and ${\rm NLL_{3}}$ give
  the improved results when the calculated $O(\alpha_s)$ contributions are also
  included, setting $m_{s}=400$ MeV, $m_{s}=500$ MeV and $m_{s}=600$ MeV,
  respectively. In this table "all" stands for the sum of all available operator 
  contributions up-to-date at the given order.}
\label{tab:o7o7BRs}
\end{table}
}        
\end{center}
\end{widetext} }

In a previous work (see Fig. 3 of Ref.~\cite{Asatrian:2015wxx}), we 
showed that the numerical impact of the self-interference contribution of
$\mathcal{O}_8$ to $\bar{B}\to X_s\gamma \gamma$ is minor in the full phase
space and no unexpected enhancements occur, therefore
it is safe to neglect this particular piece in the final numerics.

We have also investigated the relative change
    \[ A_{rel}=   \left(     \rm  \frac{   Br[\bar{B}\to X_s\gamma \gamma
      ]_{m_s-exact}^{NLL}- Br[\bar{B}\to X_s\gamma \gamma
      ]_{m_s\to0}^{NLL}  }{  \rm    Br[\bar{B}\to X_s\gamma \gamma
      ]_{m_s-exact}^{NLL} +  Br[\bar{B}\to X_s\gamma \gamma
      ]_{m_s\to0}^{NLL}   }   \right) \] 
of the NLL branching ratio 
due to the finite $m_s$ effects by
  comparing the present $m_s$ exact result with the previous approximated
  result ($m_s\to0$) of Ref. \cite{Asatrian:2014mwa}. We
  arrive at the
  following conclusion: For $m_s \in [400,\,600]$ MeV, $A_{rel}$ is 
  at most $7\%$
  when choosing the kinematical cut-off parameter $c$ as small as $1/50$. For
  larger choices of $c$, the impact on the branching ratio from terms which contain powers of $m_s$ 
  becomes even less important.

\section{Summary}\label{sec:summary}
We calculated the (${\mathcal O}_7$,\,${\mathcal O}_7$)-contribution to $\bar{B} \rightarrow X_s\gamma \gamma $ at $O(\alpha_s)$ 
 retaining the full dependence on the strange-quark mass $m_s$ in our
 results. At this order in $\alpha_s$, this requires the calculation of
 virtual corrections (with three body final state and a virtual gluon in the loop) and
 gluon bremsstrahlung corrections (tree-level contributions with four
 particles in the final state, one of them being massive).

We showed that for the phase-space region 
$(1-s_1-s_2) > c ~,{}~ ( s_1-c)\, (s_2-c) > c$  
with  $c \geq 1/50$, the branching ratio
for $\bar{B}\to X_s \gamma \gamma$ does not develop a sizable $m_s$ dependence:
the impact on the branching ratio is less than $5\%$ when $m_s$ is varied between $400$ and $600$
MeV. Besides, we have also investigated the size of the finite strange-quark
mass effects and observed that such effects are less than
$7\%$ for the same phase-space region. The observed mild sensitivity of the
branching ratio on the strange-quark mass indicates that the non-perturbative
effects related to the hadronic photon substructure are under control.

To give the complete results of our work, we append the Fortran program
``doublediff.F'' (see the corresponding paragraph after the description of Fig.~\ref{fig:DoubleWidthNLLall}
in section~\ref{sec:numerics}).

\vspace{2mm}
{\it \bf Acknowledgments}
{\small 
H.M.A. is supported by the State Committee of Science of Armenia Program Grant
No.~15T-1C161 and Volkswagen Stiftung Program Grant No. 86426. C.G. is
supported by the Swiss National Science Foundation.
A.K. acknowledges the support from the United Kingdom Science and Technology Facilities
Council (STFC) under Grant No. ST/L000431/1. We thank F. Saturnino for numerically checking the $({\mathcal O}_7,{\mathcal O}_{1,2})$
interference contributions in \eq{eq:treeRem}. A.K. is also thankful to Martin Gorbahn for numerous fruitful discussions.
}
%

\section{Appendix}\label{sec:appendix}
%

\subsection{Phase-space region for exact $m_s$ case}\label{subsec:psregion}

\begin{figure}[h]
\begin{center}
\includegraphics[width=4.0cm]{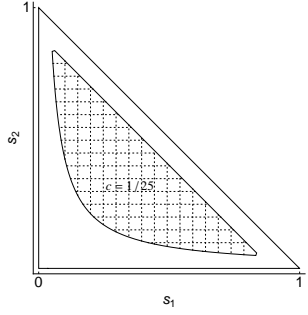}
\caption{Pictorial representation of the $(s_1,s_2)$ phase-space region when choosing $c=1/25$.}
\label{fig:phasespace}
\end{center}
\end{figure}

In this section we give the kinematical ranges considered in this
paper on the phase-space variables $s_1$ and $s_2$ in explicit form. These
restricted ranges are based on
Eqs. (\ref{eq:ps1a}) and (\ref{eq:ps2cuts}), leading to
\bea
\label{eq:PSregionExactms}
&{{\tilde s}_1}^{-} <s_{1} < {{\tilde s}_1}^{+} ~\ ; ~c  +  \frac { c }{ s_1 - c} < s_2 < 1-s_{1}-c\,\,  \\ \nn
&~~{\rm with}  \\[1mm] \nn
&{{\tilde s}_1}^{\pm}={ \left({1-c \,\pm \, \sqrt{(1-c) (1-9 c)  }} \right)/{2}  \, ,}
\eea
where $c$ is the cut-off parameter satisfying $x_4<c<1/9$. 
We display in Fig.~\ref{fig:phasespace} the geometrical representation of \eq{eq:PSregionExactms} when choosing $c=1/25$.

\subsection{Renormalization constants}\label{subsec:Renfactors}
In this appendix, we collect the  explicit expressions of the renormalization constants needed for the
ultraviolet renormalization in our calculation (see section \ref{subsec:Virtual}).

\noindent The operator ${\cal O}_{7}$, as well as the $b$- and $s$-quark mass
contained in this operator are renormalized in the $\MS$ scheme \cite{Misiak:1994zw}:
\be
 Z_{77}^{\MS} = 1 + \frac{4\,C_F}{\epsilon}\frac{\alpha_s(\mu)}{4\pi}
 + O(\alpha_s^2) \quad ; \quad
 Z_{m_b}^{\MS}=Z_{m_s}^{\MS} = 1 - \frac{3\,C_F}{\epsilon}\frac{\alpha_s(\mu)}{4\pi}
 + O(\alpha_s^2)\, .
\ee

\noindent All the remaining fields and parameters are 
renormalized in the on-shell scheme. The on-shell renormalization constants for
the $b$-quark and the $s$-quark masses read ($q=b$ or $q=s$)
\be
 Z_{m_{q}}^{\rm OS} =
 1-C_F\,\Gamma(\epsilon)\,e^{\gamma\epsilon}\,
 \frac{3-2\epsilon}{1-2\epsilon}
 \left(\frac{\mu}{m_{q}}\right)^{2\epsilon}\frac{\alpha_s(\mu)}{4\pi} +
 O(\alpha_s^2)\, ,
\ee
while the renormalization constants for the  $s$- and 
$b$-quark fields are given by ($q=b$ or $q=s$)
\bea
  Z_{2q}^{\rm OS} &=& 1 - C_F\,\Gamma(\epsilon)\,e^{\gamma\epsilon}\,\frac{3-2\epsilon}{1-2\epsilon}
  \left(\frac{\mu}{m_q}\right)^{2\epsilon}\frac{\alpha_s(\mu)}{4\pi} +
  O(\alpha_s^2)\, .
\eea

\bibliographystyle{apsrev}
\bibliography{bsgg_final}

\end{document}